\documentclass[pre,twocolumn,superscriptaddress]{revtex4}

\usepackage{amsmath}
\usepackage{amssymb}
\usepackage{graphicx}
\usepackage{color}
\usepackage[british]{babel}

\begin{document}

\title{Noisy continuous time random walks}

\author{Jae-Hyung Jeon}
\email{jae-hyung.jeon@tut.fi}
\affiliation{Department of Physics, Tampere University of Technology, FI-33101
Tampere, Finland}
\author{Eli Barkai}
\email{barkaie@mail.biu.ac.il}
\affiliation{Department of Physics, Bar Ilan University, Ramat-Gan 52900, Israel}
\author{Ralf Metzler}
\email{rmetzler@uni-potsdam.de}
\affiliation{Department of Physics, Tampere University of Technology, FI-33101
Tampere, Finland}
\affiliation{Institute for Physics \& Astronomy, University of Potsdam,
D-14476 Potsdam-Golm, Germany}
\thanks{Contribution to the special issue on Chemical Physics of Biological
Systems}

\begin{abstract}
Experimental studies of the diffusion of biomolecules in the environment of
biological cells are routinely confronted with multiple sources of stochasticity,
whose identification renders the detailed data analysis of single molecule
trajectories quite intricate. Here we consider subdiffusive continuous time
random walks, that represent a seminal model for the
anomalous diffusion of tracer particles in complex environments. This motion
is characterized by multiple trapping events with infinite mean sojourn time.
In real physical situations, however, instead of the full immobilization
predicted by the continuous time random walk model, the motion of the tracer
particle shows additional jiggling, for instance, due to thermal agitation of
the environment. We here present and analyze in detail an extension of the
continuous time random walk model. Superimposing the multiple trapping behavior
with additive Gaussian noise of variable strength, we demonstrate that the
resulting process exhibits a rich variety of apparent dynamic regimes. In
particular, such noisy continuous time random walks may appear ergodic while
the naked continuous time random walk exhibits weak ergodicity breaking.
Detailed knowledge of this behavior will be useful for the truthful physical
analysis of experimentally observed subdiffusion.
\end{abstract}

\date{\today}

\maketitle

\section{Introduction}

In a series of experiments in the Weitz lab at Harvard, Wong et al. \cite{wong}
observed that the motion of plastic tracer microbeads in a reconstituted mesh
of cross-linked actin filaments is characterized by anomalous diffusion of the
form
\begin{equation}
\label{msd}
\langle\mathbf{r}^2(t)\rangle=\int\mathbf{r}^2P(\mathbf{r},t)d\mathbf{r}\simeq
K_{\alpha}t^{\alpha}
\end{equation}
of the ensemble averaged mean squared displacement (MSD). Here, $K_{\alpha}$ is
the anomalous diffusion exponent of physical dimension $[K_{\alpha}]=\mathrm{cm}
^2/\mathrm{sec}^{\alpha}$, and the anomalous diffusion exponent $\alpha$ is in
the subdiffusive range $0<\alpha<1$. $P(\mathbf{r},t)$ is the probability
density function to find the test particle at position $\mathbf{r}$ at time $t$.
Wong et al. demonstrated that the motion
of the microbeads is represented by a random walk with subsequent immobilization
events of the beads in `cages' within the network \cite{wong}. The durations
$\tau$ of these immobilization periods were shown to follow the distribution
\begin{equation}
\psi(\tau)\sim \frac{\tau_0^\alpha}{\tau^{1+\alpha}},
\label{wtdist}
\end{equation}
with the scaling exponent $\alpha$. The exponent $\alpha$ turns out to be a
function of the ratio between the bead size and the typical mesh size. Thus,
when the bead size is larger than the mesh size, the bead becomes fully
immobilized, $\alpha=0$. Conversely, when the bead is much smaller than
the typical mesh size, it moves like a Brownian particle [$\alpha=1$, in
Eq.~\eqref{msd}], almost undisturbed by the actin mesh. However, when the bead
size is comparable to the mesh size, the motion of the bead is impeded by the
mesh, giving rise to a saltatory bead motion in between cages. While the
measured distribution (\ref{wtdist}) of immobilization times
indeed captures the statistics of the long time movement of the beads, the
measured trajectories also exhibit an additional noise, superimposed to the
jump motion according to the law (\ref{wtdist}) \cite{wong}. This additional noise
is visible as jiggling around a typical position during sojourn periods of the
bead, and appears like regular white noise.

Above example is representative for the task of physical analysis of single
particle tracking experiments. Indeed, following single particles such as
large labeled molecules, viruses, or artificial tracers in the environment
of living cells has become a standard technique in many laboratories, since
it reveals the individual trajectories without the problem of ensemble
averaging. However, the large amount of potentially noisy data also poses a
practical challenge. In physics and mathematics ideal stochastic processes
have been investigated for many years, including Brownian motion, fractional
Brownian motion, continuous time random walks, L{\'e}vy flights, etc. While
these can be reasonable approximations for the physical reality, they rarely
represent the whole story. In many cases, a superposition of at least two
types of stochastic motion are found, such as the `contamination' of the
pure stop-and-go motion described by the waiting time distribution
(\ref{wtdist}) with additional noise in the above example of the tracer beads
in the actin mesh. Similarly, Weigel et al. in their study of protein channel
motion in the walls of living human kidney cells observed a superposition of
stochastic motion governed by the law (\ref{wtdist}) and motion patterns
corresponding to diffusion on a fractal \cite{weigel}. Tabei et al. show
that the motion of insulin granules in living MIN6 cells is best explained by
a hybrid model in which fractional Brownian motion is subordinated to continuous
time random walk subdiffusion \cite{tabei}. Finally, Jeon et al. find that the
motion of lipid granules in living yeast cells is governed by continuous time
random walk motion at short times, turning over to fractional Brownian motion
at longer times \cite{lene}. Even if the trajectories are ideal,
in the lab we always have to deal with additional sources of noise, for instance,
the motion within the sample may become disturbed by a slow, random drift of the
object stage in the microscope setup, or simply by the diffusive motion on the
cover slip of a living cell, inside which the actual motion occurs that we
want to follow.

In recent years a tool box for data analysis was developed by several groups,
in particular, in the context of diffusion of tracers in the microscopic
cellular environment. These also include fundamental aspects of statistical
physics such as the ergodic properties of the underlying process and the
irreproducible nature of the time averages of the data \cite{pt}. However,
to the best of our knowledge these analysis methods and the fundamental
properties of statistical mechanics (that is, ergodicity) have not been
tested in the presence of additional noise. For instance, if we encounter
a non-ergodic process superimposed with an ergodic one, as defined precisely
below, what will we find as the result of such an experiment? 
Here we develop and study in detail a new framework for the motion of a test
particle, the \emph{noisy continuous time random walk (nCTRW)\/} subject
simultaneously to a power-law distribution (\ref{wtdist}) of sojourn times and
additional Gaussian noise. Interestingly, the noise turns
out to have a dramatic effect on the time averaged MSD, typically evaluated
in single particle tracking experiments, but a trivial effect on the ensemble
averaged MSD. We hope that this contribution will be a step towards more
realistic modeling of trajectories of single molecules, but also of other
diffusive motions of particles in complex environments.

Our analysis is based on two different scenarios for the additional Gaussian
noise. Thus, we will consider (i) a regular diffusive motion on top of the
power-law sojourn times (\ref{wtdist}). This scenario mirrors effects such
as the diffusion on the cover slip of the biological cell, in which the actual
motion occurs that we want to monitor. (ii) We study an Ornstein-Uhlenbeck
noise with a typical amplitude which may reflect an intrinsically noisy
environment, such as in the case of the tracer beads in the actin network.
We will study the MSD of the resulting motion both in
the sense of the conventional ensemble average and, for its relevance in the
analysis of single particle tracking measurement, the time average. Moreover
we demonstrate how a varying strength of the additional noise may blur the
result of stochastic diagnosis methods introduced recently, such as the scatter
of amplitudes of time averaged MSDs around the average over an ensemble of
trajectories, or the $p$-variation method.

The paper is organized as follows. In Sec.~\ref{spt}, we briefly discuss the
concepts of anomalous diffusion and single particle tracking along with the
role of (non-)er\-godicity in the context of ensemble versus time averaged MSDs.
In Sec.~\ref{analyses} we review some methods of single particle tracking
analysis. Section \ref{theory} then introduces our nCTRW model consisting of a
random walk with long
sojourn times superimposed with Gaussian noise, and we explain the simulation
scheme. The main results and discussions are presented in Secs.~\ref{diffu}
and \ref{ou}: in Sec.~\ref{diffu}, we study the statistical behavior of nCTRW
motion in the presence of superimposed free Brownian motion, while Sec.~\ref{ou}
is devoted to the motion disturbed by Ornstein-Uhlenbeck noise. Finally, the
conclusions and an outlook are presented in Sec.~\ref{conc}. In the Appendix,
we provide a brief outline of the stochastic analysis tools used to quantify
the nCTRW processes studied in Secs.~\ref{diffu} and \ref{ou}. Note that in the
following, for simplicity we concentrate on the one-dimensional case; all
results can easily be generalized to higher dimensions.

\section{Anomalous diffusion and single particle tracking}

In this Section we briefly review the continuous time random walk (CTRW) model
and the behavior of CTRW time series. Moreover, we present some common methods
to analyze traces obtained from single particle tracking experiments or
simulations.

\subsection{Continuous time random walks and time averaged mean squared
displacement}
\label{spt}

Anomalous diffusion of the subdiffusive form (\ref{msd}) with $0<\alpha<1$ is
quite commonly observed in a large variety of systems and on many different
scales \cite{havlin,bouchaud,report,report1}. Examples range from microscopic
systems such as the motion of small tracer particles in living biological cells
and similar crowded systems \cite{golding,hoefling,saxton1}, over the motion of charge
carriers in amorphous semiconductors \cite{scher} to the dispersion of chemical
tracers in groundwater systems \cite{grl}.

A prominent model to describe the subdiffusion law (\ref{msd}) is given by the
Montroll-Weiss-Scher continuous time random walk (CTRW) \cite{montroll,scher}.
Originally applied to describe extensive data from the stochastic motion of
charge carriers in amorphous semiconductors \cite{scher}, the CTRW model finds
applications in many areas. \emph{Inter alia}, CTRW subdiffusion was shown to
underlie the motion of microbeads in reconstituted actin networks \cite{wong},
the subdiffusion of lipid granules in living yeast cells \cite{lene,natascha}
of protein channels in human kidney cells \cite{weigel}, and insulin granules
in MIN6 cells \cite{tabei}, as well as the temporal
spreading of tracer chemicals in groundwater aquifers \cite{grl,brian}. While
the sojourn times in a subdiffusive CTRW follow the law (\ref{wtdist}), the
length of individual jumps is governed by a distribution $\lambda(x)$ of jump
lengths, with finite variance $\sigma^2=\int x^2\lambda(x)dx$. In the simplest
case of a jump process on a lattice, each jump is of the length of the lattice
constant. Physically, the power-law form (\ref{wtdist}) of sojourn times may
arise due to multiple trapping events in a quenched energy landscape with
exponentially distributed depths of traps \cite{monthus,burov}. Subdiffusive
CTRWs macroscopically exhibit long-tailed memory effects, as characterized by
the dynamic equation for the probability density function $P(x,t)$ containing
fractional differential operators, see below \cite{mebakla}.

Usually, we think in terms of the ensemble average (\ref{msd}) when we talk
about the MSD of a diffusive process. However, starting with Ivar Nordlund's
seminal study of the Brownian motion of a slowly sedimenting mercury droplet
\cite{nordlund} and now routinely performed even on the level of single
molecules \cite{wong,golding,lene,natascha,weigel,tabei,weber,sm,bronstein,
szymansky,natascha1,caspi,banks}, 
the time series $x(t)$ obtained from measuring the
trajectory of an individual particle is evaluated in terms of the \emph{time
averaged\/} MSD
\begin{equation} 
\overline{\delta^2(\Delta)}=\frac{1}{T-\Delta}\int_0^{T-\Delta}\Big(x(t+\Delta)
-x(t)\Big)^2dt
\label{tamsd}
\end{equation}
where $\Delta$ is the so-called lag time, and $T$ is the overall measurement
time. For Brownian motion, the time averaged MSD (\ref{tamsd}) is equivalent
to the ensemble averaged MSD (\ref{msd}), $\langle x^2(\Delta)\rangle=\overline{
\delta^2(\Delta)}$, if only the measurement time $T$ is sufficiently long to
ensure self-averaging of the process along the trajectory. This fact is a
direct consequence of the ergodic nature of Brownian motion \cite{pt,stas2}.
To obtain reliable behaviors for the time averaged MSD for trajectories with
finite $T$ one often introduces the average
\begin{equation}
\label{eatamsd}
\left<\overline{\delta^2(\Delta)}\right>=\frac{1}{N}\sum_{i=1}^N\overline{
\delta^2_i(\Delta)}
\end{equation}
over several individual trajectories $\overline{\delta^2_i(\Delta)}$.

What happens when the process is anomalous, of the form (\ref{msd})? There
exist anomalous diffusion processes, that are ergodic in the above sense that
for sufficiently long $T$ we observe the equivalence $\langle x^2(\Delta)
\rangle=\overline{\delta^2(\Delta)}$. Prominent examples are given by diffusion
on random fractal geometries \cite{yasmin}, as well as fractional
Brownian motion and the related fractional Langevin equation motion for which
ergodicity is approached algebraically \cite{deng,igorg,jae}. Such behavior was in
fact corroborated in particle tracking experiments in crowded environments
\cite{szymansky,bronstein,natascha1} and simulations of lipid bilayer systems
\cite{kneller,matti}. In superdiffusive L{\'e}vy walks time and ensemble
averaged MSDs
differ merely by a factor \cite{lws,zukla,froemberg}. However, the diverging
sojourn time in the
subdiffusive CTRW processes naturally causes weak ergodicity breaking
in the sense that even for extremely long measurement times $T$ ensemble and
time averages no longer coincide \cite{bouchaud1,barkai,zaid}. For the time
averaged MSD the consequences are far-reaching. Thus, for subdiffusive CTRW we
find the somewhat counterintuitive result that the time averaged MSD for free
motion scales \emph{linearly\/} with the lag time, $\langle\overline{\delta^2(
\Delta,T)}\rangle\simeq K_{\alpha}\Delta/T^{1-\alpha}$ for $\Delta\ll T$, and
thus
$\langle x^2(\Delta)\rangle\neq\overline{\delta^2(\Delta)}$ \cite{He,lubelski}.
In particular, the result $\langle\overline{\delta^2(\Delta)}\rangle$ decreases
with the measurement time $T$, an effect of ageing. Concurrently, the MSD of
subdiffusive CTRWs depends on the time difference between the preparation of the
system and   the start of the measurement \cite{johannes}. Under confinement,
while the
ensemble averaged MSD saturates towards the thermal value, the time averaged
MSD continues to grow as $\langle\overline{\delta^2(\Delta)}\rangle\simeq\Delta^
{1-\alpha}$ as long as $\Delta\ll T$ \cite{stas1,neusius}. Another important
consequence of the weakly non-ergodic behavior is that the time averaged
MSD (\ref{tamsd}) remains irreproducible even in the limit of $T\rightarrow
\infty$: the amplitude of individual time averaged MSD curves $\overline{\delta
^2_i(\Delta)}$ scatters significantly between different trajectories, albeit
with a well-defined distribution \cite{He,igor}. In other words, this scatter
of amplitudes corresponds to a distribution of diffusion constants
\cite{saxton2}. Active transport of molecules in the cell exhibits
superdiffusion with non-reproducible results for the time averages \cite{caspi2},
a case treated theoretically only recently \cite{lws,froemberg,akimoto}.

To pin down a given stochastic mechanism underlying some measured trajectories,
several diagnosis methods have been developed. Thus, one may analyze the first
passage behavior \cite{olivier}, moment ratios and the statistics of mean maximal
excursions \cite{vincent}, the velocity autocorrelation \cite{stas2,matti}, the
statistics of the apparent diffusivities \cite{radons}, or the $p$-variation of
the data \cite{marcin,kepten}. In the following we analyze the sensitivity of the
time averaged MSD and its amplitude scatter as well as the $p$-variation method
to noise, that is superimposed to naked subdiffusive CTRWs. In particular, we
show that at larger amplitudes of the additional noise the CTRW-inherent weak
non-ergodicity may become completely masked.

\subsection{A primer on single particle trajectory analysis}
\label{analyses}

Before proceeding to define the nCTRW process, we present a brief review of
several techniques developed recently for the analysis of single particle
trajectories.

\begin{figure*}
\begin{center}
\includegraphics[width=5.8cm]{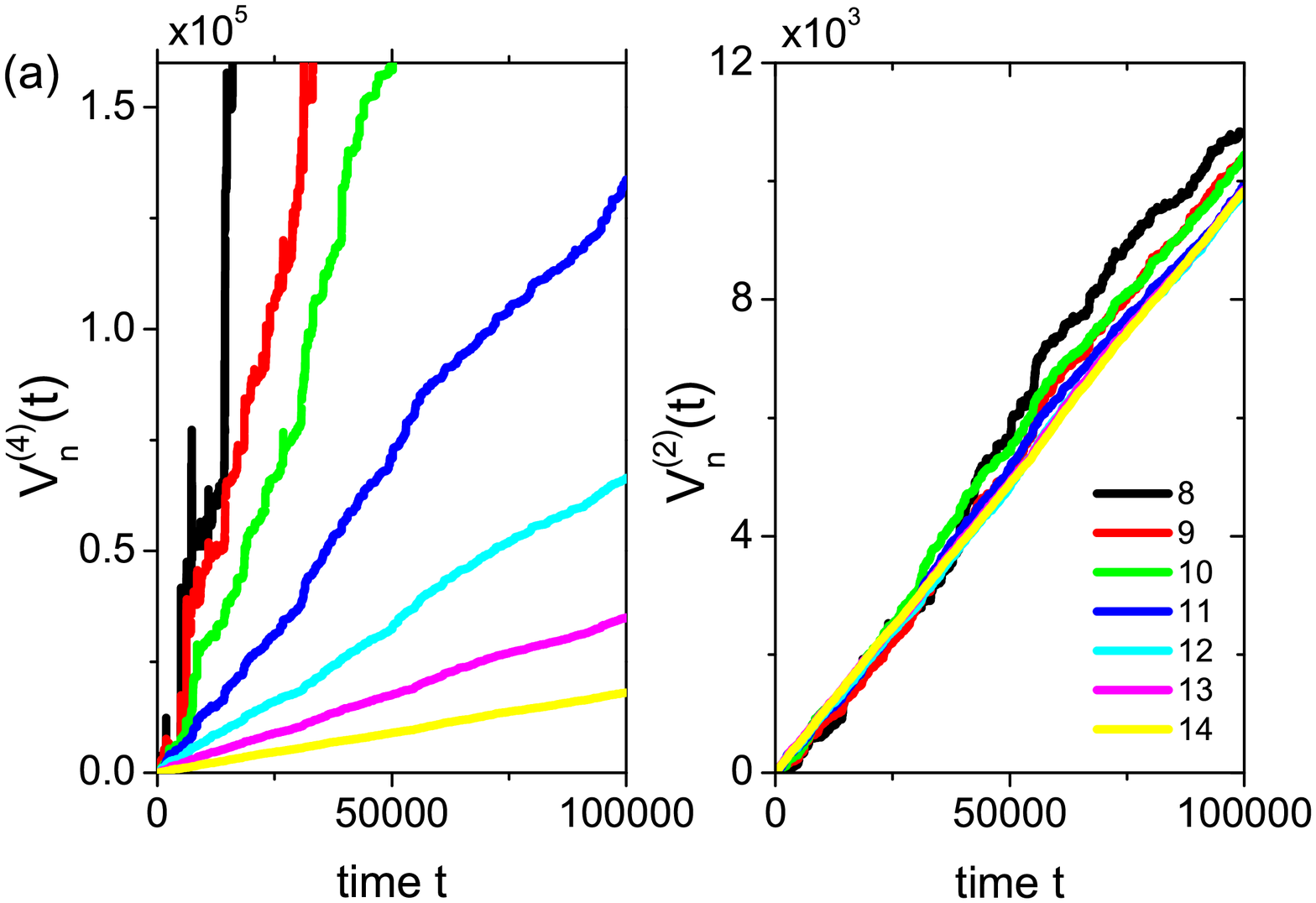}
\includegraphics[width=5.8cm]{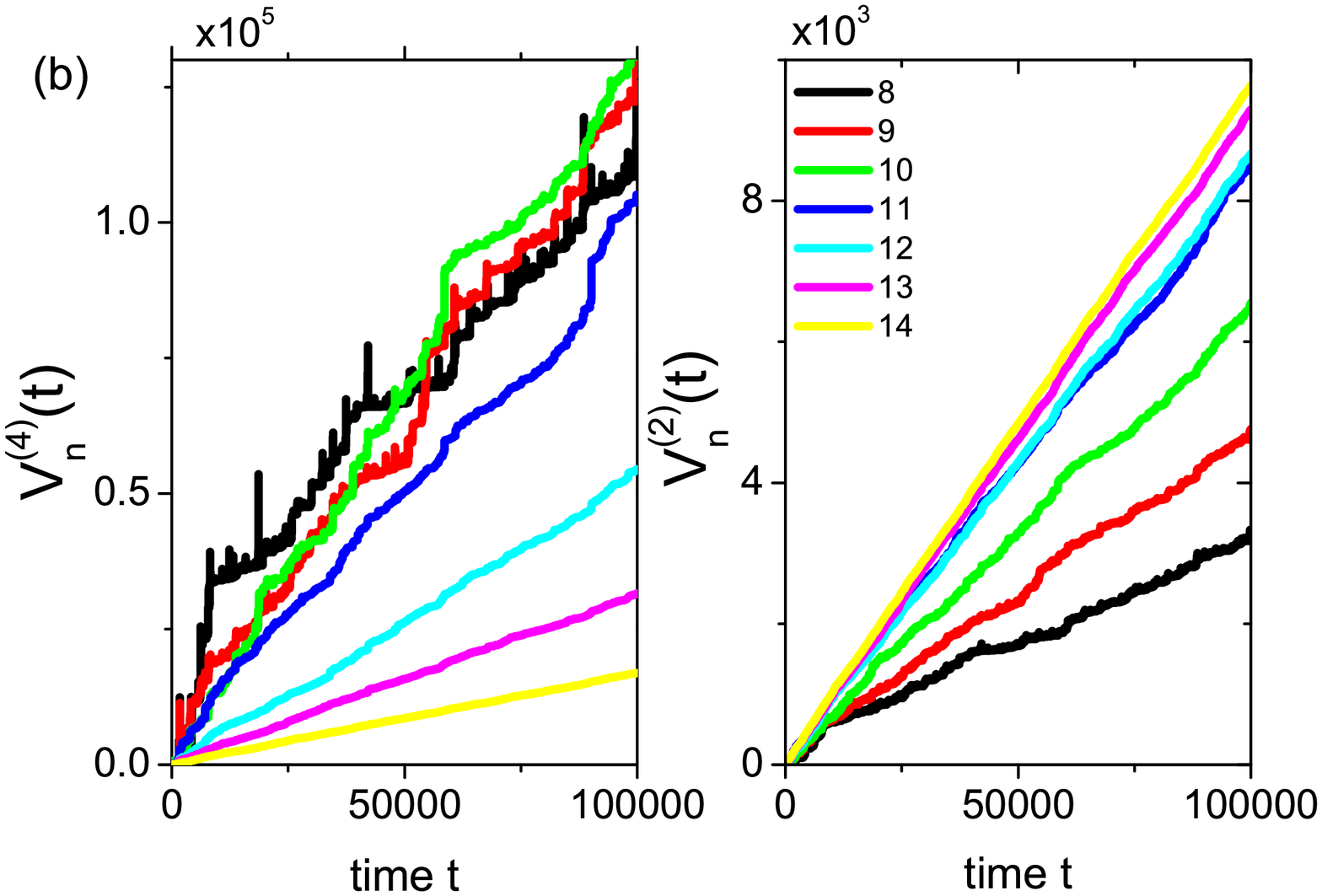}
\includegraphics[width=5.8cm]{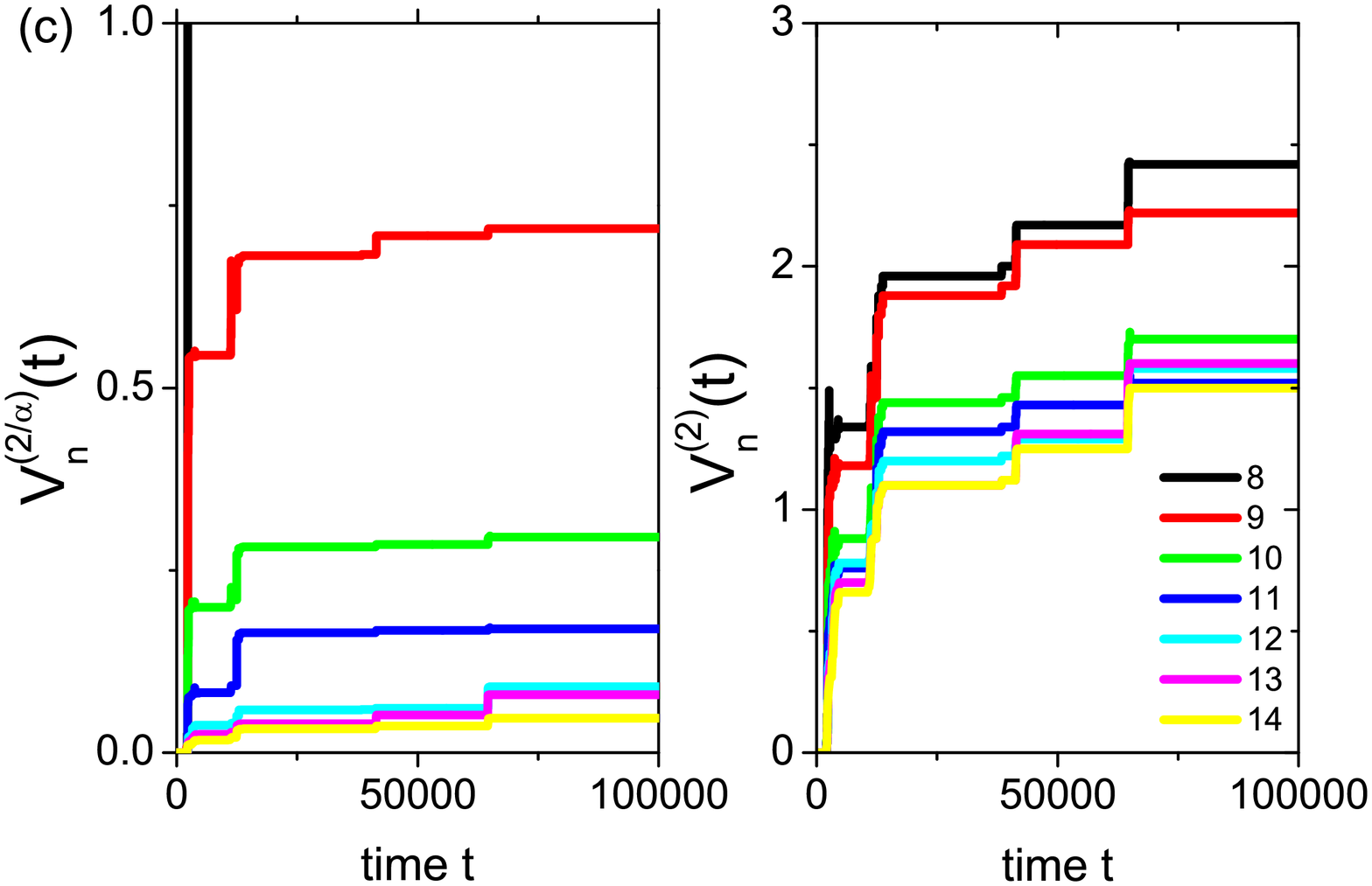}
\caption{Results of $p$-variation test for (a) Brownian noise $z_B(t)$, (b)
Ornstein-Uhlenbeck noise $z_{OU}(t)$, and (c) subdiffusive CTRW $x_\alpha(t)$
with $\alpha=0.5$. The left panels in each case are for $p=4$ for both (a) and
(b), and (c) $p=2/\alpha=4$. All right panels are for $p=2$. The color coding
refers to the values of $n$ indicated in the right panel of each pair of graphs.}
\label{pvar}
\end{center}
\end{figure*}

\subsubsection{Mean squared displacement}

Already from the time averaged MSD $\overline{\delta^2_i}$ of individual
particle traces important information on the nature of the process may be
extracted, provided that the measurement time $T$ is sufficiently long. Thus,
for ergodic processes the time and ensemble averaged MSD are equivalent,
$\langle\overline{\delta^2(\Delta)}\rangle=\langle x^2(\Delta)\rangle$. A
significant difference between both quantities points at non-ergodic behavior.
In particular, the scatter of amplitudes (distribution of diffusion constants)
between individual traces $\overline{
\delta^2_i}$ turns out to be a quite reliable measure for the (non-)ergodicity
of a process. Thus, in terms of the dimensionless parameter $\xi=\overline{
\delta^2}/\langle\overline{\delta^2}\rangle$ the distribution of amplitudes
$\phi(\xi)$ has a Gaussian profile centered on the ergodic value $\xi=1$ for
finite-time ergodic processes \cite{jeon}. Its width narrows with increasing
$T$, eventually approaching the sharp distribution $\delta(\xi-1)$ at $T\to
\infty$, i.e., each trajectory gives exactly the same result, equivalent to
the ensemble averaged MSD. For subdiffusive CTRW processes $\phi(\xi)$ is
quite broad and has a finite value at $\xi=0$ for completely stalled
trajectories \cite{stas2,He,jeon,johannes,heinsalu}. For $\alpha\le1/2$ the
maximum of the
distribution is at $\xi=0$, while for $\alpha>1/2$ the maximum is located at
$\xi=1$. The non-Gaussian distribution $\phi(\xi)$ for subdiffusive CTRWs is
almost independent of $T$, and its shape is already well established even for
relatively short trajectories \cite{jeon}.

\subsubsection{$p$-variation test}

The $p$-variation test was recently promoted as a tool to distinguish CTRW
and FBM-type subdiffusion \cite{marcin}. It is defined in terms of the sum of
increments of the trajectory $x(t)$ on the interval $[0,T]$ as
\begin{equation}
V_n^{(p)}(t)=\sum_{j=0}^{2^n-1}\left|x\left(\frac{(j+1)T}{2^n}\wedge t\right)-x
\left(\frac{jT}{2^n}\wedge t\right)\right|^p,
\label{psum}
\end{equation}
where $a\wedge b=\mathrm{min}\{a,b\}$. The quantity $V^{(p)}=\lim_{n\to\infty}
V_n^{(p)}$ has distinct properties for certain stochastic processes. Thus, for
both free and confined Brownian motion $V^{(2)}(t)\sim t$ and $V^{(p)}(t)=0$ for any
$p>2$. For fractional Brownian motion, $V^{(2)}(t)=\infty$,
while $V^{(2/\alpha)}(t)\sim t$. Finally, for CTRW subdiffusion, $V^{(2)}(t)$ features
a step-like, monotonic increase as function of time $t$, and $V^{(2/\alpha)}(t)=0$.

Fig.~\ref{pvar} shows the $p$-variation results for Brownian noise $z_B(t)$,
Ornstein-Uhlenbeck noise $z_{OU}(t)$, and the naked subdiffusive CTRW $x_{
\alpha}(t)$. We plot the sum (\ref{psum}) for various, finite values of $n$.
For the Brownian noise $z_B(t)$, $V_n^{(4)}$ monotonically decreases with
growing $n$, a signature of the predicted convergence to $V^{(4)}\rightarrow0$.
The sum $V^{(2)}$ appears independent of $n$ and proportional to time $t$, as
expected.
For the Ornstein-Uhlenbeck noise $z_{OU}(t)$ we observe that $V^{(4)}$ scales
linearly with $t$ and the slope decreases with growing $n$, indicating a
convergence to zero. Also $V^{(2)}$ is linear in $t$. For smaller $n$, the
slope increases with $n$ and saturates at large $n$.
Finally, for CTRW subdiffusion $V^{(2/\alpha)}$ appears to converge to zero
for increasing $n$, as predicted. In contrast, $V^{(2)}$ has the distinct,
monotonic step-like increase expected for CTRW subdiffusion. A more detailed
description of the $p$-variation for Ornstein-Uhlenbeck noise and CTRW
subdiffusion is found in Appendix \ref{app_pvar}.

\section{Noisy continuous time random walk}
\label{theory}

In this Section we define the nCTRW model and describe our simulations scheme
used in the following Sections.

\subsection{Two nCTRW models}

We consider an nCTRW process $x(t)$ in which ordinary CTRW subdiffusion $x_{
\alpha}(t)$ with anomalous diffusion exponent $0<\alpha<1$ is superimposed with
the Gaussian noise $\eta z(t)$,
\begin{equation}
\label{superimdef}
x(t)=x_\alpha(t)+\eta z(t).
\end{equation}
That means that the Gaussian process $z(t)$ is additive and thus independent of
$x_\alpha(t)$. The relative strength of the additional Gaussian noise is
controlled by the amplitude parameter $\eta\geq0$. In this study we consider
the following two Gaussian processes: (1) in the first case $z(t)$ is a simple
Brownian diffusive process $z_B(t)$ with zero mean $\langle z_B(t)\rangle=0$ and
variance $\langle z_B^2(t)\rangle=2Dt$. As mentioned, physically this could
represent the (slow) diffusion of a living bacteria or endothelial cell on the
cover slip while we want to record the motion of a tracer inside the cell, or
the random drifting of the experimental
stage. (2) In the second case $z(t)$ represents Ornstein-Uhlenbeck process
$z_{OU}(t)$, the confined Brownian motion in an harmonic potential (see
Eq.~\eqref{oun} for the definition of the process). Stochastic processes of
this second kind, for example, likely mimic the motion of a bead confined in a
polymer network or immersed in a macromolecularly crowded environment, where
the thermal agitation of the confining environment gives rise to the random
fluctuation of the position of the trapped tracer particle around some average
value.

According to the definition \eqref{superimdef}, nCTRW possesses have the
following generic properties. Its probability density function (PDF) $P(x,t)$
is given by the convolution of the individual PDFs $P_{\alpha}(x,t)$ of a CTRW
subdiffusion process and $P_G(x,t)$ of the Gaussian process,
\begin{equation}
P(x,t)=\int_{-\infty}^{\infty}P_{\alpha}(x-y,t)P_G(y,t)dy.
\label{nctrwsol}
\end{equation}
This chain rule states that a given position $x$ of the combined process is
given by the product of the probability that the CTRW process has reached the
position $x-y$ and the Gaussian process contributes the distance $y$, or vice
versa. Here
the PDF $P_{\alpha}(x,t)$ satisfies the fractional Fokker-Planck equation
\cite{report}
\begin{equation}
\frac{\partial}{\partial t}P_{\alpha}(x,t)={_0}\mathcal{D}_t^{1-\alpha}K_{
\alpha}\frac{\partial^2}{\partial x^2}P_{\alpha}(x,t)\label{FFPE}
\end{equation}
where the Riemann-Liouville fractional derivative of $P(x,t)$ is
\begin{equation}
_0\mathcal{D}_t^{1-\alpha}P_{\alpha}(x,t)=\frac{1}{\Gamma(\alpha)}\frac{
\partial}{\partial t}\int_0^t\frac{P_{\alpha}(x,t')}{(t-t')^{1-\alpha}}dt'.
\end{equation}
Physically, this fractional operator thus represents a memory integral with
a slowly decaying kernel. From definition (\ref{superimdef}) it follows that
the ensemble averaged MSD is given by
\begin{eqnarray}
\nonumber
\langle x^2(t)\rangle&=&\int_{-\infty}^{\infty}x^2P_\alpha(x,t)dx+\int_{-\infty}
^{\infty} y^2P_G(y,t)dy\\
&=&\langle x_\alpha^2(t)\rangle+\eta^2\langle z^2(t)\rangle.
\label{eamsd}
\end{eqnarray}
The characteristic function of $P(x,t)$ according to
Eq.~(\ref{nctrwsol}) is given by the product
\begin{equation} 
\label{charnctrw}
P(q,t)=\int_{-\infty}^{\infty}e^{iqx}P(x,t)dx=P_{\alpha}(q,t)P_G(q,t)
\end{equation}
of the characteristic functions of the individual processes, $P_{\alpha}$ and
$P_G$. We here use the simplified notation that the Fourier transform of a
function is expressed by its explicit dependence on the Fourier variable $q$.
With the Mittag-Leffler function $E_\alpha(x)=\sum_{m=0}^{\infty}x^m/\Gamma(1+
\alpha m)$ we find that \cite{report}
\begin{equation}
P_{\alpha}(q,t)=E_\alpha\left(-q^2K_{\alpha}t^\alpha\right),
\end{equation}
assuming that the CTRW process starts at $t=0$ with initial conditions $x_{
\alpha}(0)=0$. The characteristic function $P_{\alpha}(q,t)$ initially decays
like a stretched exponential $P_{\alpha}(q,t)\approx\exp(-q^2K_\alpha t^\alpha
/\Gamma(1+\alpha))$ and has the asymptotic power-law decay $P_{\alpha}(q,t)\sim
1/(q^2K_\alpha t^\alpha)$.

The Brownian noise $\eta z_B(t)$ with initial condition $z_B(0)=0$ has the
characteristic function
\begin{equation}
P_G(q,t)=\exp\left(-\eta^2Dq^2t\right).
\label{charB}
\end{equation}    
In the nCTRW process this Brownian noise always dominates the dynamics of the
process at long times, since the exponential relaxation \eqref{charB} dominates
the characteristic function $P(q,t)$. The Ornstein-Uhlenbeck noise $\eta z_{OU}
(t)$ with $z_{OU}(0)=0$, defined in Eq.~\eqref{oun}, has the characteristic function
\begin{equation}
P_G(q,t)=\exp\left(-\frac{\eta^2Dq^2}{2k}\left(1-e^{-2kt}\right)\right).
\label{charOU}
\end{equation}
At short times $t\ll k^{-1}$ when confinement by the harmonic Ornstein-Uhlenbeck
potential is negligible, Eq.~\eqref{charOU} reduces to the result \eqref{charB}
for Brownian noise, and thus the characteristic functions of the two nCTRW
processes are identical. At longer times the characteristic function
\eqref{charOU} saturates to $P_G(q)=\exp[-\eta^2Dq^2/(2k)]$. This means that
the long-time behavior of the nCTRW with superimposed Ornstein-Uhlenbeck noise,
largely reflects the naked CTRW process $x_\alpha(t)$, if the noise level
is not too high. Keeping these general features in mind, we further study
the statistical quantities of the two nCTRW processes numerically.

\subsection{Simulation of the nCTRW process}

To simulate the nCTRW process we independently obtain time traces of the
subdiffusive CTRW motion $x_{\alpha}(t)$ and the additional Gaussian noise.
The motion $x_{\alpha}(t)$ with $0<\alpha<1$ is generated on a lattice of
spacing $a$ from the normalized waiting time distribution
\begin{equation}
\label{wtd1}
\psi(\tau)=\frac{\alpha/\tau_0}{(1+\tau/\tau_0)^{1+\alpha}}
\end{equation}
with the power-law asymptotic scaling $\psi(\tau)\sim\alpha\tau_0^{\alpha}/\tau
^{1+\alpha}$. Here $\tau_0$ is a scaling constant of dimension $[\tau_0]=
\mathrm{sec}$.
The jump lengths are determined by the $\delta$-distribution $\lambda(x)=\frac{
1}{2}\delta(|x|-a)$ \cite{lene,jae}. Within this construction the CTRW process
is associated with the fractional Fokker-Planck equation \eqref{FFPE} with the
anomalous diffusion exponent \cite{eli}
\begin{equation}
K_\alpha=\frac{\Gamma(1-\alpha)a^2}{2\tau_0^\alpha}.
\label{Kalpha}
\end{equation} 
In the simulations we choose $\tau_0=1$,
and consequently in the following times are given in units of $\tau_0$, which
is also chosen equal to the time increments $\delta t$, at which the system is
updated. The lattice spacing is $a=0.1$. We simulate the cases $\alpha=
0.5$ and $0.8$.

\begin{figure*}
\centering
\includegraphics[width=8cm,angle=0]{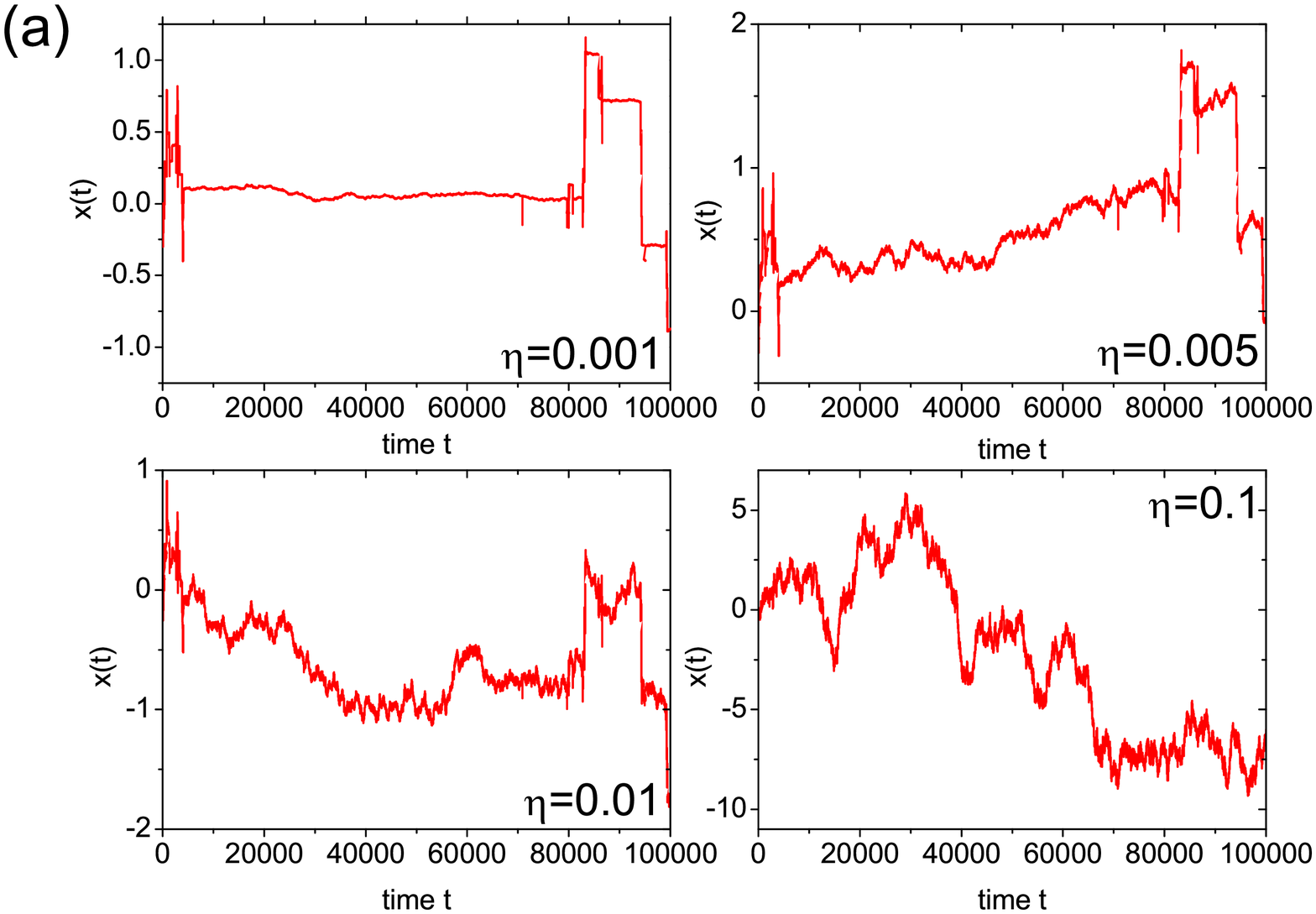}
\includegraphics[width=8cm,angle=0]{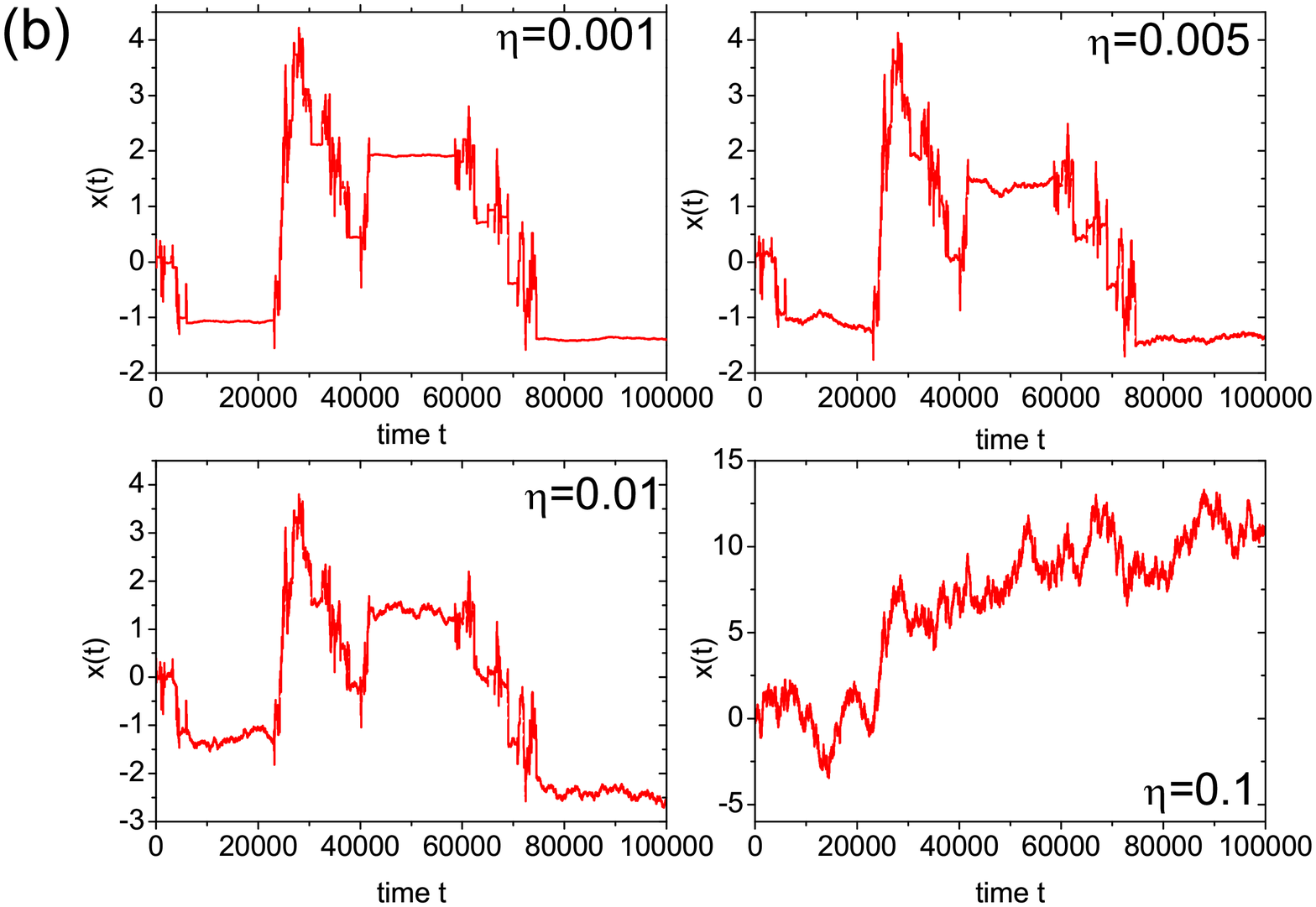}
\caption{Sample trajectories of the nCTRW process $x_\alpha(t)$ with
superimposed Brownian noise $\eta z_B(t)$ of strength $\eta$, for several
values of $\eta$, and for anomalous diffusion exponents (a) $\alpha=0.5$
and (b) $\alpha=0.8$. With increasing noise strength the otherwise pronounced
sojourn states become increasingly blurred by the Brownian motion. For each
$\alpha$ the CTRW part of the trajectory is identical.}
\label{bmtra05}
\end{figure*}

The added Gaussian noise is obtained as follows. Brownian noise $z(t)=z_B(t)$
is obtained at discrete times $t_n=n\delta t$ (with $\delta t=1$), in terms of
the Brownian walk
\begin{equation}
z_B(t_n)=\sum_{k=1}^{n}\delta t\sqrt{2D}\xi_B(t_k) ,
\end{equation}
where $\xi_B(t)$ represents white Gaussian noise of zero mean and unit variance
$1/\delta t$ with our choice $\delta t=1$. In the simulations we take $D=0.05$,
such that $\langle z_B^2(1)\rangle=0.1$. The Ornstein-Uhlenbeck noise $z(t)=
z_{OU}(t)$ is obtained by integration of the Langevin equation
\begin{equation}
\label{oun}
\frac{d}{dt}z_{OU}(t)=-kz_{OU}(t)+\sqrt{2D}\xi_{B}(t)
\end{equation}
with the initial condition $z_{OU}(0)=0$. Here, the coefficient of the restoring
force is chosen as $k=0.01$. For both Brownian and Ornstein-Uhlenbeck noise,
the following values for the noise strength are used: $\eta=0.001$, $0.005$,
$0.01$, and $0.1$.

\section{nCTRW with Brownian noise}
\label{diffu}

In Fig.~\ref{bmtra05} we show typical examples of simulated trajectories of the
nCTRW process $x(t)$ with added Brownian noise. The trajectories $x(t)$ for
different noise strengths $\eta$ is constructed for the \emph{same\/} CTRW
process $x_\alpha(t)$, i.e., only the noise strength varies in between the
panels. This way it is easier to appreciate the influence of the added noise.
We generally observe that the trajectories preserve the profile of the
underlying CTRW process $x_\alpha(t)$ at small $\eta$, and they become quite
distorted from the original CTRW trajectory $x_{\alpha}(t)$ for larger values
of $\eta$. In particular, for the largest noise strength the character of the
pure CTRW with its pronounced stalling events is completely lost, and visually
one might judge these traces to be pure Brownian motion, at least when the
length $T$ of the time series is not too large. Moreover, the effect of the
noise is stronger for smaller $\alpha$. This is because long stalling events
occur more frequently as $\alpha$ decreases, and thus the actual displacement
of the process is also smaller and the influence of the Brownian motion becomes
relatively more pronounced.

\subsection{Ensemble-averaged mean squared displacements}

\begin{figure}
\includegraphics[width=8.8cm,angle=0]{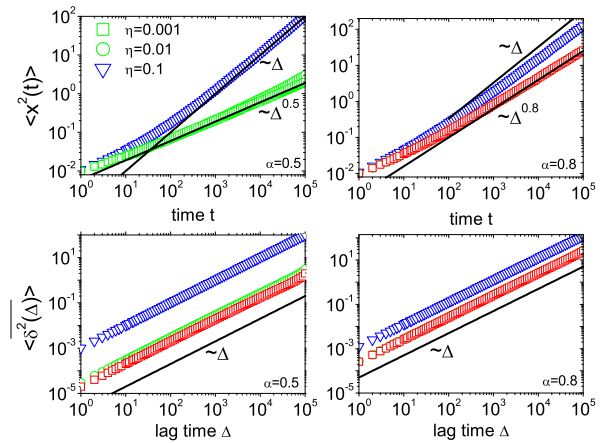}
\caption{Top: Ensemble averaged MSD of the nCTRW process $x(t)$ with noise
strengths $\eta=0.001$, $0.01$, and $0.1$, for anomalous diffusion exponents
$\alpha=0.5$ (left) and $\alpha=0.8$ (right). A turnover from subdiffusive
to Brownian (linear) scaling is observed. Bottom: trajectory-averaged time
averaged MSD $\langle\overline{\delta^2(\Delta)}\rangle$ for $x(t)$ for the
same values of $\eta$ and $\alpha$. The overall measurement time is $T=10^5$
in units of $\delta t$.}
\label{bmtamsd}
\label{bmeamsd}
\end{figure}

\begin{figure*}
\begin{center}
\includegraphics[width=8cm,angle=0]{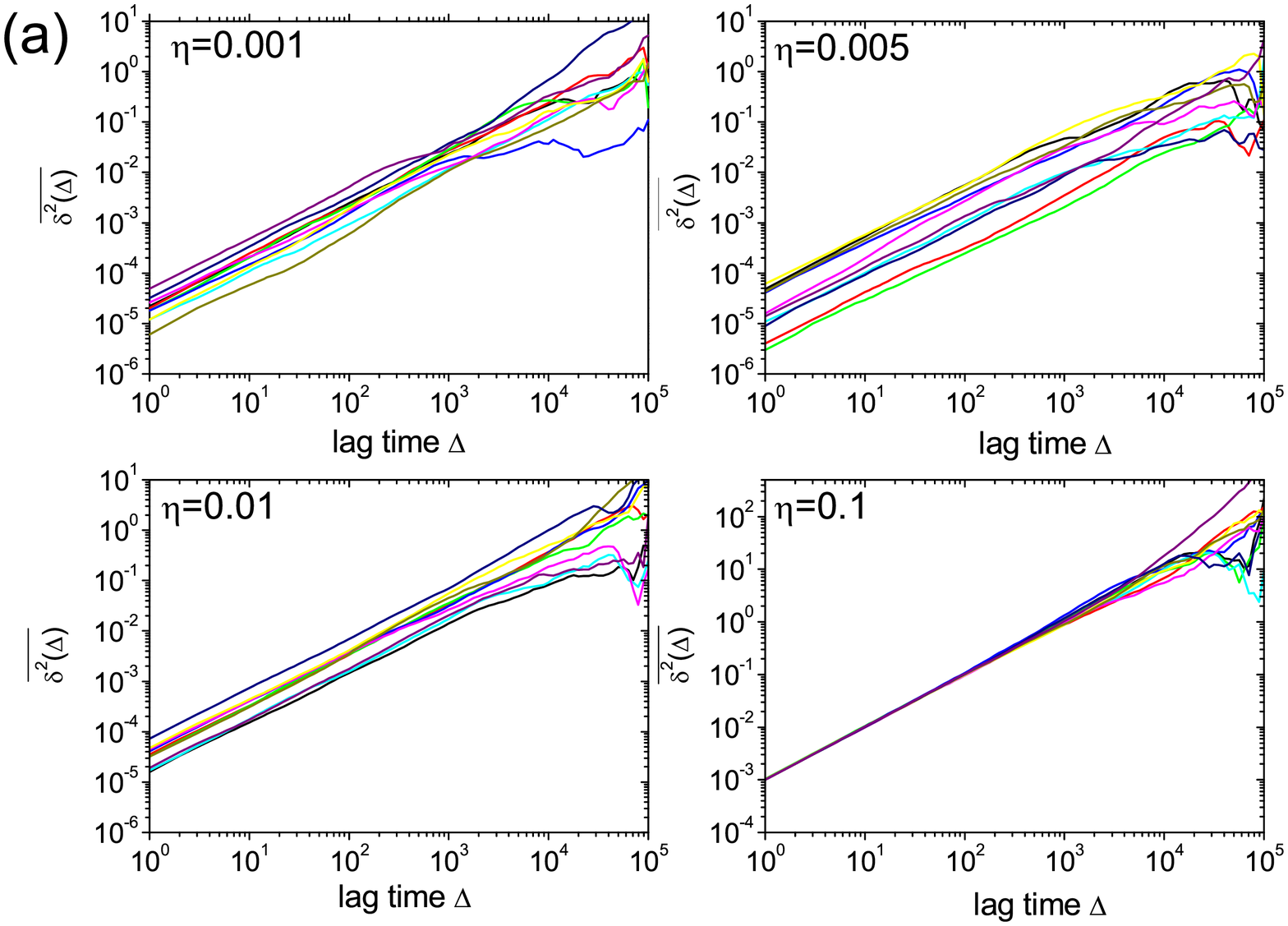}
\includegraphics[width=8cm,angle=0]{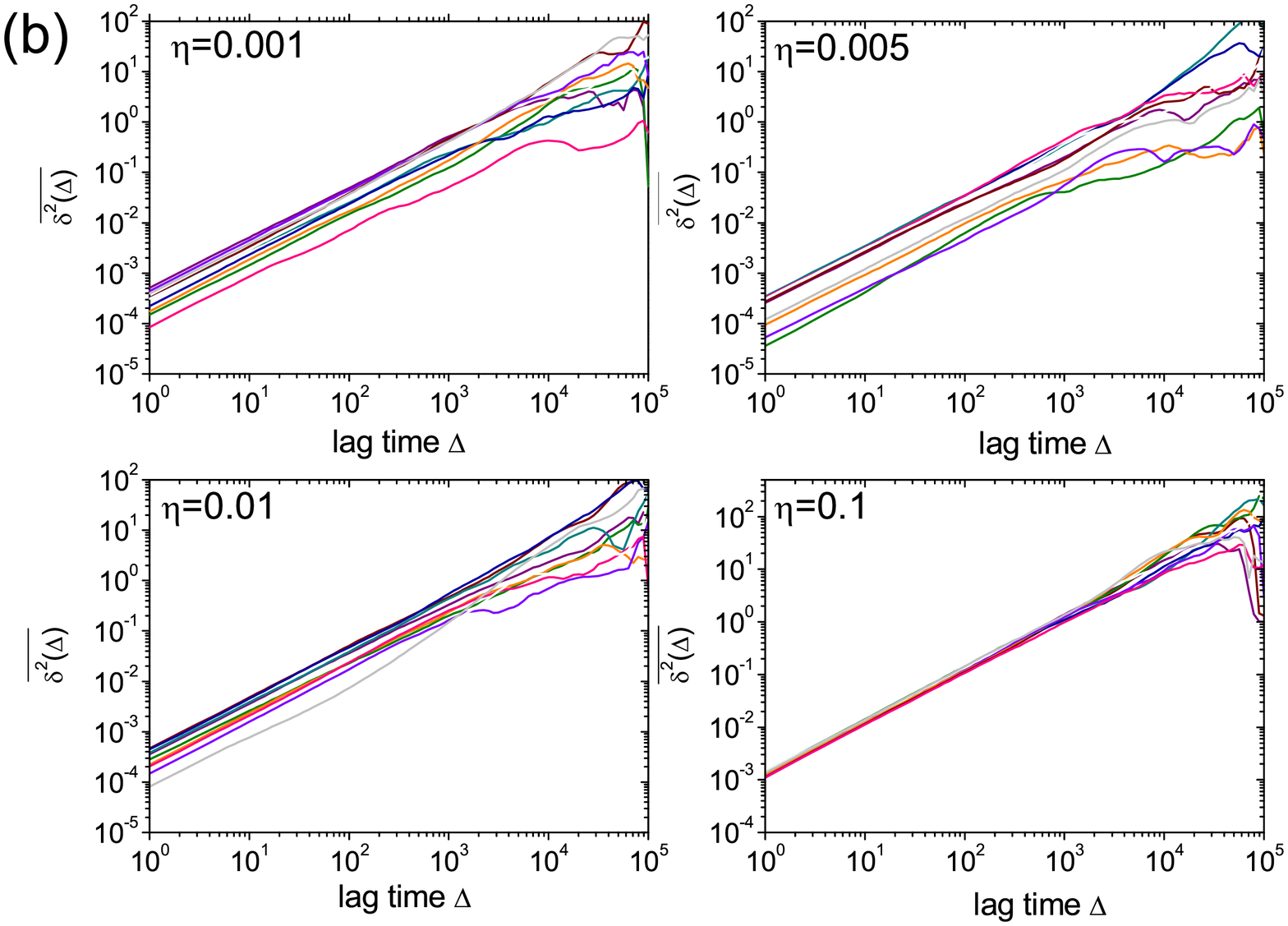}
\caption{Ten individual time averaged MSD curves for nCTRW with (a) $\alpha=0.5$
and (b) $\alpha=0.8$ with added Brownian
noise, for the same parameters as in Fig.~\ref{bmtamsd}. The relative amplitude
scatter dramatically reduces at high noise strength.}
\label{bmtamsds}
\end{center}
\end{figure*}

From the simulated trajectories we evaluate the ensemble-averaged MSD for the
nCTRW and study how its scaling behavior is affected by the Brownian noise $z_B
(t)$. Fig.~\ref{bmeamsd} summarizes the results for the nCTRW process with
$\alpha=0.5$ and $\alpha=0.8$. In both cases, a common feature is that the
ensemble averaged MSD exhibits a continuous transition from subdiffusion with
anomalous diffusion exponent $\alpha$ to normal diffusion. This occurs either as
the noise strength $\eta$ is increased at fixed time $t$, or as time $t$ is
increased at a fixed $\eta$. Due to the additivity of the two contributions we
obtain
\begin{equation}
\langle x^2(t)\rangle=\frac{2K_{\alpha}}{\Gamma(1+\alpha)}t^\alpha+2\eta^2Dt,
\label{bmmsd}
\end{equation} 
which is valid as long as $t$ is considerably larger than the time increment
$\delta t$. Eq.~(\ref{bmmsd}) demonstrates that for the subdiffusive CTRW
processes $x_{\alpha}(t)$ with $0<\alpha<1$ the ensemble averaged MSD of the
nCTRW $x(t)$ has a crossover in its scaling from $\simeq t^\alpha$ at short times
to $\simeq t$ at long times, with the crossover time scale $t_c\sim\left(K_{
\alpha}/[\Gamma(1+\alpha)\eta^2D]\right)^{1/(1-\alpha)}$. That is, the effect
of the Brownian
noise $z_B(t)$ emerges only at long times $t>t_c$. Conversely, below $t_c$ the
process appears to behave as the bare CTRW process. Note that the crossover time
$t_c$ rapidly decreases with increasing $\eta$ as $t_c\sim\eta^{-2/(1-\alpha)}$.
This explains why we only observe normal diffusion behavior without crossover
for the largest noise strength $\eta=0.1$. The crossover time $t_c$ also rapidly
increases as the exponent $\alpha$ approaches to one [as $K_\alpha\sim\Gamma(1-
\alpha)$ in our choice of $\tau_0=1$, see Eq.~\eqref{Kalpha}]. Accordingly, in
Fig.~\ref{bmeamsd} the ensemble averaged MSDs for $\alpha=0.8$ do not fully
reach the linear regime within the time window of our simulation, in contrast
to the case for $\alpha=0.5$ with $\eta=0.1$.

\subsection{Time-averaged mean squared displacement}

We now consider the individual time averaged MSDs $\overline{\delta^2(\Delta)}$
of the nCTRW process from single trajectories $x(t)$ according to our definition
in Eq.~(\ref{tamsd}). Fig.~\ref{bmtamsd} presents the trajectory-to-trajectory
average (\ref{eatamsd}). In all cases we find that the time averaged MSDs grow
linearly with lag time $\Delta$ in the entire range of $\Delta$, showing a clear
disparity from the scaling behavior of the ensemble averaged MSDs above. In the
time averaged MSD the Brownian noise $z_B(t)$ simply affects the apparent
diffusion constant, that is, the effective amplitude of the linear curves. To
understand this phenomenon quantitatively, we obtain the analytical form of the
trajectory-averaged time averaged MSD,
\begin{eqnarray}
\left<\overline{\delta^2(\Delta)}\right>\sim\frac{2K_\alpha\Delta}{\Gamma(1+
\alpha)T^{1-\alpha}}+2\eta^2D\Delta,
\label{bmeatamsd}
\end{eqnarray}
valid for $\Delta\ll T$. Eq.~\eqref{bmeatamsd} shows that both contributions
the naked CTRW $x_\alpha(t)$ and the Brownian process $\eta z_B(t)$ are
linearly proportional to the lag time $\Delta$. Note that the exponent $\alpha$
and the noise strength $\eta$ only enter into the apparent diffusion constant
\begin{equation}
D_{\mathrm{app}}\equiv\frac{K_{\alpha}T^{\alpha-1}}{\Gamma(1+\alpha)}+\eta^2D,
\end{equation}
where $\langle\overline{\delta^2(\Delta)}\rangle=2D_{\mathrm{app}}\Delta$. This
result indicates that the effect of the Brownian noise cannot be noticed when we
exclusively consider the scaling behavior of the time averaged MSD. Also in
terms of the apparent diffusion constant, one hardly notices the presence of the
Brownian noise component as long as $\eta$ is small. This agrees with our
observations for the time averaged MSD curves in Fig.~\ref{bmtamsd} for noise
strengths $\eta=0.001$ and $0.01$.

\begin{figure}
\centering
\includegraphics[width=8.8cm,angle=0]{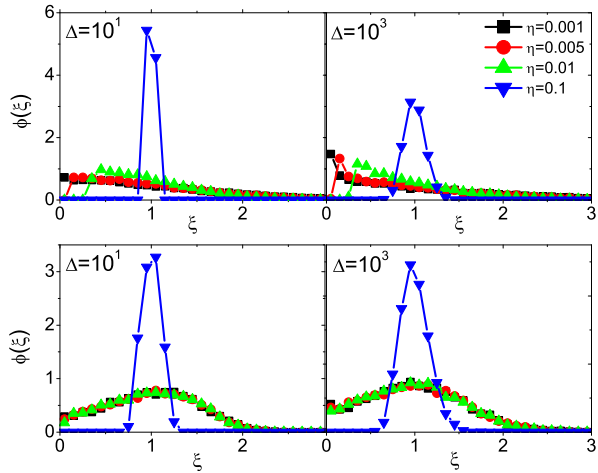}
\caption{Variation of normalized
scatter distributions $\phi(\xi)$ as a function of dimensionless
variable $\xi=\overline{\delta^2}/\langle\overline{\delta^2}\rangle$
for  the cases of $\eta=0.001$ (black square), 0.005 (red circle), 0.01
(green upper-triangle), and 0.1 (blue down-triangle). (Upper panels)
$\alpha=0.5$. (Lower panels) $\alpha=0.8$. In each figure the results
were obtained from $10^4$ runs. }
\label{bmscatter}
\end{figure}

\begin{figure*}
\centering
\includegraphics[width=12cm,angle=0]{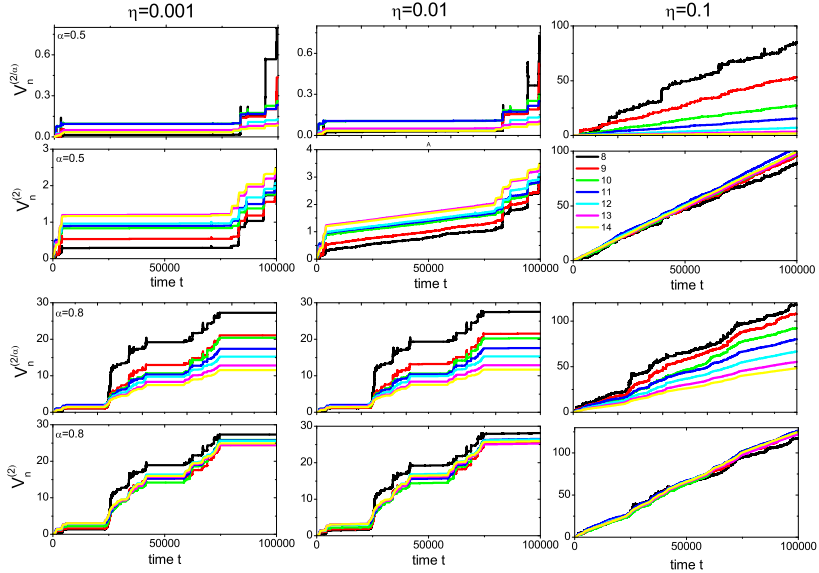} 
\caption{Results of the $p$-variation test for the nCTRW process with Brownian
noise $\eta z_{B}(t)$ for noise strengths $\eta=0.001$,
$\eta=0.01$, and $\eta=0.1$. The upper (lower) two rows are for
$\alpha=0.5$ and $0.8$. In all figures the $p$ sums are plotted with the same
color code: $n=8$ (black), 9 (red), 10 (green), 11 (blue), 12 (cyan), 13
(violet), and 14 (yellow).}
\label{pvar1}
\end{figure*}

We also check the fluctuations between individual time averaged MSD curves
$\overline{\delta^2(\Delta)}$. Each panel in Fig.~\ref{bmtamsds} plots ten
individual time averaged MSDs for the nCTRW process. In all cases, the
individual time averaged MSDs display linear scaling with lag time $\Delta$,
namely, the scaling behavior of $\langle\overline{\delta^2(\Delta)}\rangle$.
The individual amplitudes scatter, that is, the apparent diffusion constant
$D_{\mathrm{app}}$ fluctuates. With increasing strength of the Brownian
component the relative scatter between individual trajectories dramatically
diminishes, leading to an apparently ergodic behavior. Thus, for the largest
noise strength $\eta=0.1$, the ten trajectories almost fully collapse onto a
single curve for $\Delta\ll T$.

\subsection{Scatter distribution}

We quantify the amplitude scatter of the individual time averaged MSDs in terms
of the normalized scatter distributions $\phi(\xi)$, where the dimensionless
variable $\xi$ stands for the ratio $\xi=\overline{\delta^2}/\langle\overline{
\delta^2}\rangle$ of individual traces $\overline{\delta^2}$ versus the
trajectory average $\langle\overline{\delta^2}\rangle$ (compare the derivations
in Refs.~\cite{pt,He,stas2}). Fig.~\ref{bmscatter} shows $\phi(\xi)$ for several
values of the lag time $\Delta$. When the noise is negligible ($\eta=0.001$),
the observed broad distribution is nearly that of the pure CTRW process. For
$\alpha=1/2$ the distribution has the expected Gaussian profile centered at
$\xi=0$ \cite{He}, indicating that long stalling events of the order of the
entire measurement time $T$ occur with appreciable probability. In the opposite
case $\eta=0.1$, the Brownian noise results in a relatively sharply peaked,
bell-shaped distribution typical for ergodic processes, at all lag times. An
interesting effect of the Brownian noise is that at intermediate strengths it
only tends to suppress the contribution at around zero while it does not
significantly change the overall profile of the distribution compared to the
noise-free case. This implicates that the trajectories share non-ergodic and
ergodic elements. For instance, the trajectory of the nCTRW process $x(t)$
itself for $\eta=0.01$ in Fig.~\ref{bmtra05} shows a substantially blurred
profile of the underlying CTRW process $x_\alpha(t)$ due to the relatively
strong
Brownian noise; however, the time averaged MSD and its distribution do exhibit
non-ergodic behavior, as shown in Fig.~\ref{bmscatter}. The case $\alpha=0.8$
shown in Fig.~\ref{bmscatter} features similar albeit less pronounced
effects. In particular, $\phi(\xi)$ for the naked CTRW has its maximum at the
ergodic value $\xi=1$. Again, the effect of the Brownian noise is to suppress
the contribution at $\xi=0$.

\subsection{$p$-variation test}

\begin{figure*}
\includegraphics[width=8cm,angle=0]{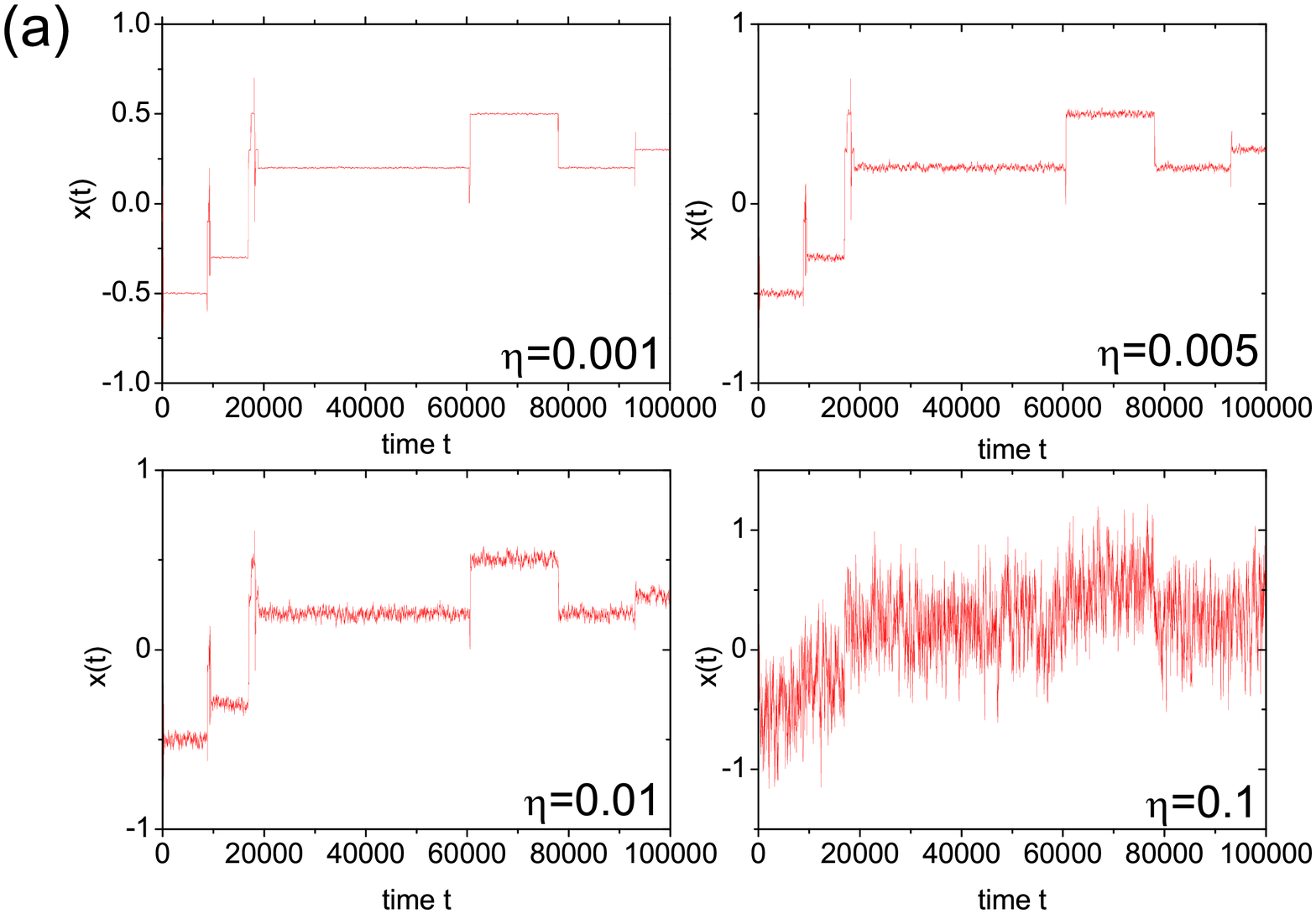}
\includegraphics[width=8cm,angle=0]{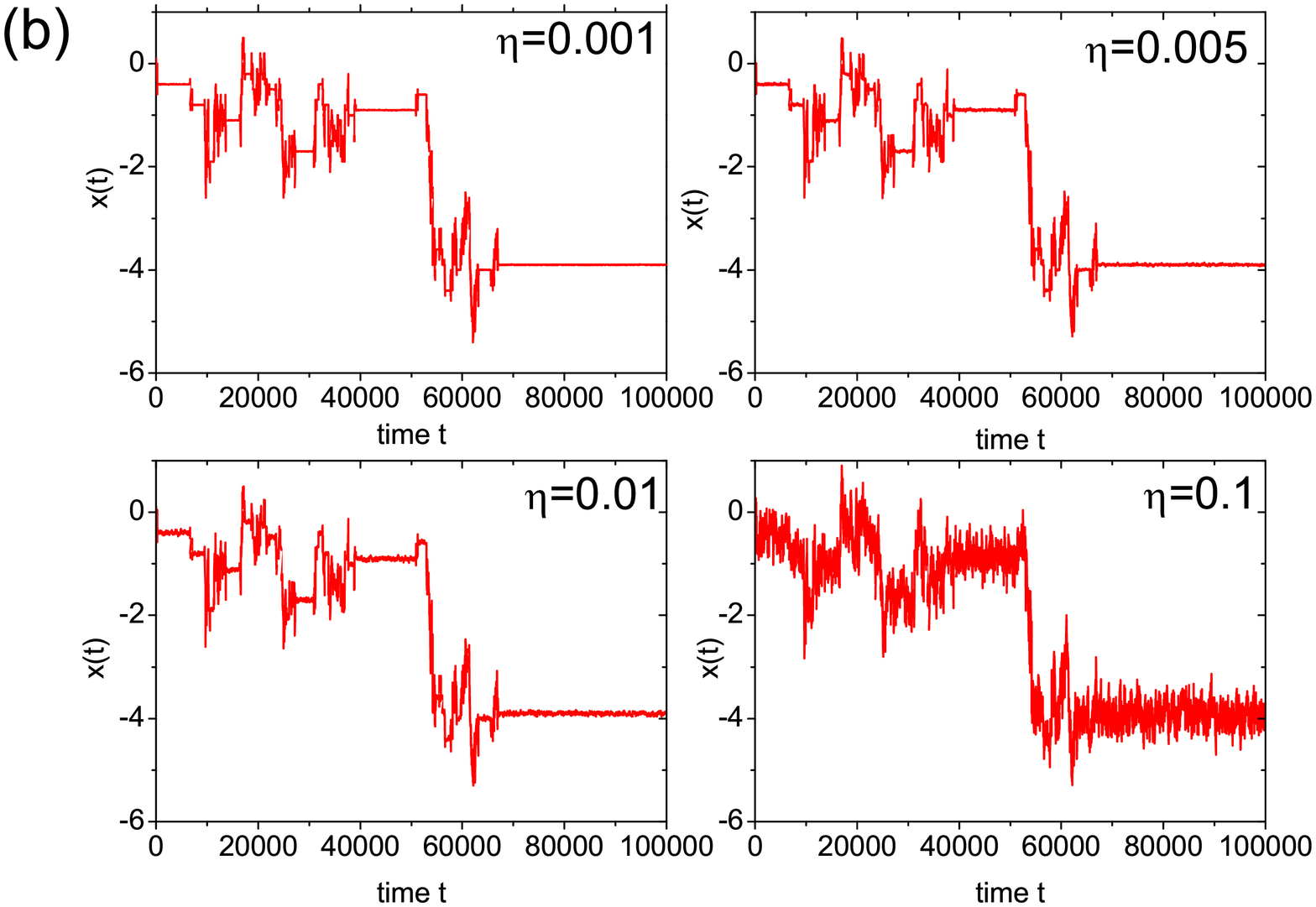}
\caption{Sample trajectories of the nCTRW process $x(t)$ with Ornstein-Uhlenbeck
noise for several values of the noise strength $\eta$ and anomalous diffusion
exponents (a) $\alpha=0.5$ and (b) $\alpha=0.8$. In contrast to the Brownian
noise case of Fig.~\ref{bmtra05}, however, the approximately constant amplitude
of the superimposed noise is characteristic for the Ornstein-Uhlenbeck process.}
\label{outra}
\end{figure*}

We now turn to the $p$-variation test and investigate its sensitivity to the
additional noise in the nCTRW process, in comparison to the established
results for the naked CTRW $x_\alpha(t)$. From the trajectory $x(t)$, the
partial sum $V_n^{(p)}(t)$ is calculated for finite $n$ according to definition
(\ref{psum}), where we choose $p=2/\alpha$ and $p=2$, compare Section
\ref{analyses}. In Fig.~\ref{pvar1}, we plot the results of the $p$-variation
at increasing $n$ for the nCTRW process. We observe that for both cases
$\alpha=0.5$ and $0.8$, the $p$-sums behave analogously to the predictions for
the naked CTRW process, as long as the noise strength remains sufficiently
small, according to Fig.~\ref{pvar1} this holds for $\eta=0.001$ and 0.005 (not
shown). In
this case, $V_n^{(2/\alpha)}(t)$ monotonically decreases with increasing $n$,
indicating the limiting behavior $V_n^{(2/\alpha)}(t)\rightarrow0$ for large $n$.
Meanwhile, $V_n^{(2)}(t)$ approaches the monotonic step-like behavior typical
for the CTRW process $x_\alpha(t)$, as $n$ increases. Note that the $p$-sums
have plateaus in their increments in analogy to the noise-free case due
to the long stalling events in the trajectory.

As the magnitude of the Brownian noise grows larger ($\eta=0.01$ and
0.1), however, the behavior of the $p$-sums changes significantly. While
$V_n^{(2/\alpha)}(t)$ decreases with increasing $n$ as for the weaker noise
case, it increases linearly with $t$ nearly without any sign of plateaus
for both nCTRW processes of $\alpha=0.5$ and 0.8 when the noise strength
is increased to $\eta=0.1$. This new feature is the expected behavior of
$V_n^{(p)}(t)$ with $p>2$ for a Brownian diffusive process (see
Sec.~\ref{analyses} and Fig.~\ref{pvar}).
Indeed, it can be shown that for $p=4$ the $p$-sum of the
Brownian noise $z_B(t)$ behaves as $V_n^{(4)}(t)\sim(\frac{T}{2^n})t$. We note
that for the Brownian noise $z_B(t)$ the $p$-sum $V_n^{(p)}(t)$ with
$p>2$ always decays out to zero as $n\rightarrow\infty$. Therefore,
the nCTRW process will always have the same $p$ variation result of
$V_n^{(2/\alpha)}=0$ (in the limit of $n\rightarrow\infty$) as the pure CTRW
$x_\alpha(t)$, even in case that its profile is dominated by large noise.

In contrast, the $p$-sum $V_n^{(2)}(t)$ exhibits a more distinguished effect
of the Brownian noise due to the fact that the Brownian noise $\eta z_B(t)$ has
$V_n^{(2)}(t)\simeq \eta^2 2Dt$. Especially, we find that the noise effect is
pronounced for the nCTRW with $\alpha=0.5$, when the underlying CTRW process
$x_\alpha(t)$ features only few jumps. In this case, the Brownian noise of
moderate strength ($\eta=0.01$) causes an incline with almost identical slope
to the step-like profiles of $V_n^{(2)}(t)$. At the strongest noise $\eta=0.1$,
the step-like behavior typical for the naked CTRW process $x_\alpha(t)$ is
nearly masked and the overall tendency follows that of Brownian motion shown
in Fig.~\ref{pvar}. Accordingly, the $p$-variation test does not properly pin
down the underlying CTRW process $x_\alpha(t)$ and thus potentially produces
inconsistent conclusion for the nCTRW process. A qualitatively identical
behavior is obtained for the nCTRW process when the underlying CTRW process
$x_\alpha(t)$ performs relatively frequent jumps (corresponding to the case
of $\alpha=0.8$). However, here the effect of the Brownian noise appears weak,
because the contribution of $z_B(t)$ relative to the magnitude of $x_\alpha(t)$ becomes smaller at larger $\alpha$ values, see the trajectories $x(t)$ in
Fig.~\ref{bmtra05}.

\section{nCTRW with Ornstein-Uhlenbeck noise}
\label{ou}

We now turn to nCTRW processes $x(t)$, in which the superimposed noise is of
Ornstein-Uhlenbeck form (\ref{oun}). In this case the influence of the added
noise is expected to diminish as the process develops, according to our
discussion in Sec.~\ref{theory}.

Fig.~\ref{outra} shows simulated trajectories for the nCTRW process $x_\alpha
(t)$ with two different anomalous diffusion exponents, (a) $\alpha=0.5$ and (b)
$\alpha=0.8$, for different noise strengths $\eta$. Indeed we observe that,
compared to case of Brownian noise depicted in Fig.~\ref{bmtra05} the profiles
of the jumps and rests of the naked CTRW motion $x_\alpha(t)$ are relatively
well preserved despite the mixing with the Ornstein-Uhlenbeck noise $z_{OU}(t)$.
Note that the simulated trajectories with moderate noise strength appear quite
similar to the experimental traces of the microbeads in the reconstituted actin
network reported by Wong et al.~\cite{wong}.
The noise interference appears considerably lesser for the case $\alpha=0.8$,
when the magnitude of the net displacements is large relative to the noise
contribution due to frequent jumps in the trajectory $x_\alpha(t)$ when
$\alpha$ is closer to unity. Only for the highest noise strength the pure CTRW
behavior with its distinct sojourns becomes blurred by the Ornstein-Uhlenbeck
noise.

\begin{figure}
\includegraphics[width=8.8cm,angle=0]{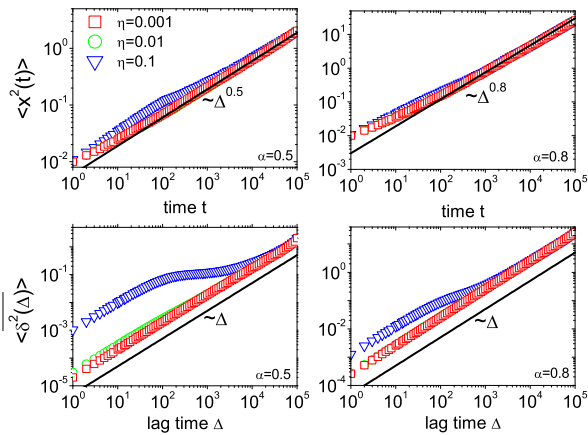}
\caption{Top: Ensemble-averaged MSD $\langle x^2(t)\rangle$ of the nCTRW
process with Ornstein-Uhlenbeck noise for anomalous diffusion exponents
$\alpha=0.5$ and $0.8$. Bottom: Trajectory-average of the time
averaged MSD $\langle\overline{\delta^2(\Delta)}\rangle$ for the same
$\alpha$. We use $T=10^5$.}
\label{fig:outamsd}
\label{oueamsd}
\end{figure}

\subsection{Ensemble-averaged mean squared displacement} 

In Fig.~\ref{oueamsd} we plot the ensemble averaged MSD $\langle x^2(t)\rangle$
of the nCTRW process with Ornstein-Uhlenbeck noise for different noise strengths
$\eta$. We note that, regardless of the intensity $\eta$, the ensemble averaged
MSDs follow the scaling law $\sim t^{\alpha}$ of the noise-free CTRW process
$x_\alpha(t)$, in particular, at long times. Moreover, all MSD curves at
different $\eta$ almost collapse onto each other, although small differences
are discernible at short times. These results suggest that, in contrast to the
Brownian noise case discussed in Sec.~\ref{diffu}, the Ornstein-Uhlenbeck noise
does not critically interfere with the diffusive behavior of the noise-free CTRW
motion, as expected. To obtain a quantitative understanding of these results we
derive the analytic form of the ensemble averaged MSD,
\begin{equation}
\langle x^2(t)\rangle=\frac{2K_\alpha}{\Gamma(1+\alpha)}t^\alpha+\frac{\eta^2
D}{k}\left(1-e^{-2kt}\right).
\label{oumsd}
\end{equation}
Here the last term stems from the contribution of the Ornstein-Uhlenbeck noise.
For $t>k^{-1}$ it saturates to the constant $\eta^2 D/k$, which is typically
small relative to $2K_\alpha t^\alpha/\Gamma(1+\alpha)$. Hence, at long times
$t>k^{-1}$, the
noise term is negligible, and the ensemble averaged MSD grows as $2K_\alpha t^{
\alpha}/\Gamma(1+\alpha)$,
consistent with the observations in Fig.~\ref{oueamsd}. In the
opposite case for $t<k^{-1}$, the contribution of the noise is expanded to
obtain $\langle z_{OU}^2(t)\rangle\approx2\eta^2Dt$. As discussed for the
characteristic function \eqref{charOU}, on these time scales the
Ornstein-Uhlenbeck noise $z_{OU}(t)$ has the same form as the Brownian noise
$z_B(t)$, leading to the same scaling form (\ref{bmmsd}) for the ensemble
averaged MSD. However, as studied in the previous case, the effect of the
linear scaling is irrelevant at short times.

\subsection{Time-averaged mean squared displacement}

We now turn to the time averaged MSD curves $\overline{\delta^2(\Delta)}$ from
individual trajectories $x(t)$ of the nCTRW process. Fig.~\ref{fig:outamsd}
shows the trajectory-averaged time averaged MSD $\langle\overline{\delta^2(
\Delta)}\rangle$ for different noise strengths. For both anomalous diffusion
exponents $\alpha=0.5$ and $0.8$, we observe qualitatively the same behavior.
On the one hand, the CTRW motion superimposed with moderate noise ($\eta=0.001$
and $0.01$) leads to a linear scaling of the time averaged MSD with lag
time $\Delta$, with almost identical amplitude. On the other hand, when the
noise amplitude becomes large, the scaling of the time averaged MSD is
significantly affected. We find that these results are consistent with the
analytical form of the trajectory-averaged time averaged MSD,
\begin{eqnarray}
\left<\overline{\delta^2(\Delta)}\right>\sim\frac{2K_\alpha\Delta}{\Gamma(1+
\alpha)T^{1-\alpha}}+\frac{2\eta^2 D}{k}\left(1-e^{-k\Delta}\right),
\label{outamsd}
\end{eqnarray}
valid at lag times $\Delta\ll T$. In this expression it is worthwhile to point
out that the contribution of the naked CTRW process $x_\alpha(t)$ is decreased
as the length of the trajectory becomes longer, due to the aging effect of the
decreasing effective diffusion constant $\simeq T^{\alpha-1}$, while the noise
is independent of $T$. Due to this effect the time averaged MSD (\ref{outamsd})
has three distinct scaling regimes: (i) at lag times $\Delta\gg k^{-1}$,
the time averaged MSD is linearly proportional to the lag time with apparent
diffusion constant $D_{\mathrm{app}}\approx K_\alpha T^{\alpha-1}/\Gamma(1+
\alpha)$. (ii)
At lag times $\Delta\ll k^{-1}$, the time averaged MSD is again proportional
to $\Delta$. On this timescale, however, the noise part cannot be ignored,
and the apparent diffusion constant is given by $D_{\mathrm{app}}\approx
K_\alpha T^{\alpha-1}/\Gamma(1+\alpha)+\eta^2D$. Note that the diffusion
constant at short lag
times is larger than the one at long times. (iii) For lag times $\Delta\approx
k^{-1}$, the time averaged MSD is that of confined Brownian diffusion where
$\langle \overline{\delta^2(\Delta)}\rangle\approx2K_\alpha T^{\alpha-1}\Delta/
\Gamma(1+\alpha)
+2\eta^2D$. These three regimes are expected to occur only in the presence of
large noise strengths when $\eta^2D>K_\alpha T^{\alpha-1}/\Gamma(1+\alpha)$
(see the case
$\eta=0.1$ in Fig.~\ref{fig:outamsd}). In the opposite case, when $2\eta^2D$
is negligible compared to $K_\alpha T^{\alpha-1}/\Gamma(1+\alpha)$, the three
regimes are
indistinguishable, and only one scaling law $\langle\overline{\delta^2(\Delta)}
\rangle\approx2 K_\alpha T^{\alpha-1}\Delta/\Gamma(1+\alpha)$ is observed at
all lag times. This
behavior is seen for the two cases of $\eta=0.001$ and $0.01$ in
Fig.~\ref{fig:outamsd}.

\begin{figure}
\centering
\includegraphics[width=8.8cm,angle=0]{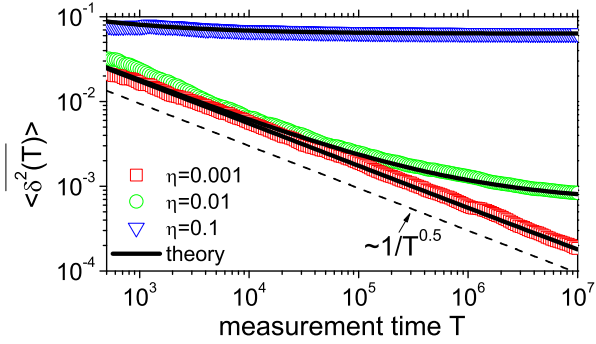}
\caption{Profiles of the time averaged MSD $\langle\overline{\delta^2(\Delta,T)}
\rangle$ as a function of measurement time $T$ at a fixed lag time $\Delta=100$.
We show the nCTRW of $\alpha=0.5$ with $\eta z_{OU}(t)$ at $\eta=0.001$, 0.01,
and 0.1 (from bottom to top). Solid and dotted lines represent the analytical
form \eqref{outamsd} and the scaling $\sim T^{\alpha-1}$, respectively.}
\label{fig:agingou}
\end{figure}

\begin{figure*}
\includegraphics[width=8cm,angle=0]{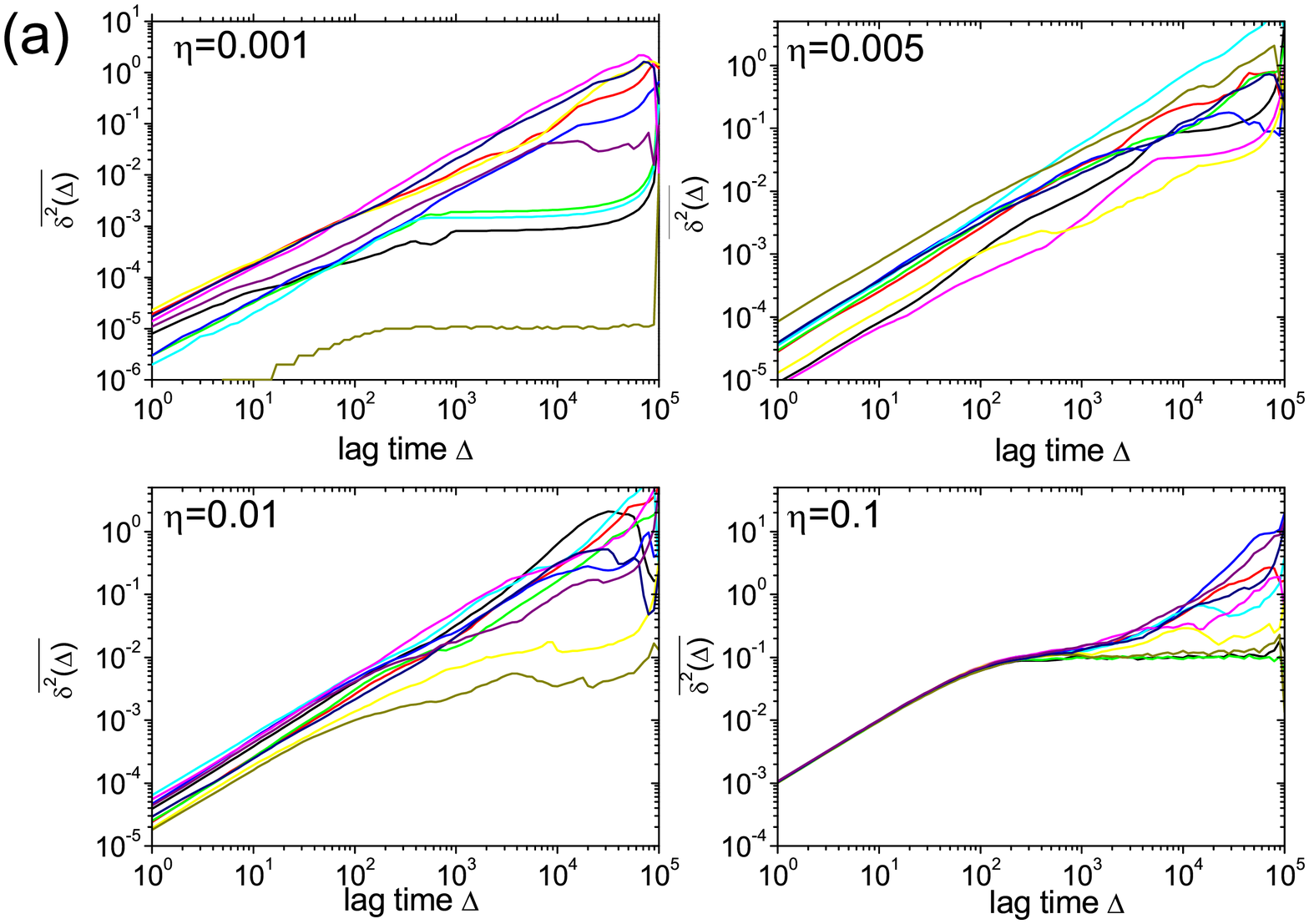}
\includegraphics[width=8cm,angle=0]{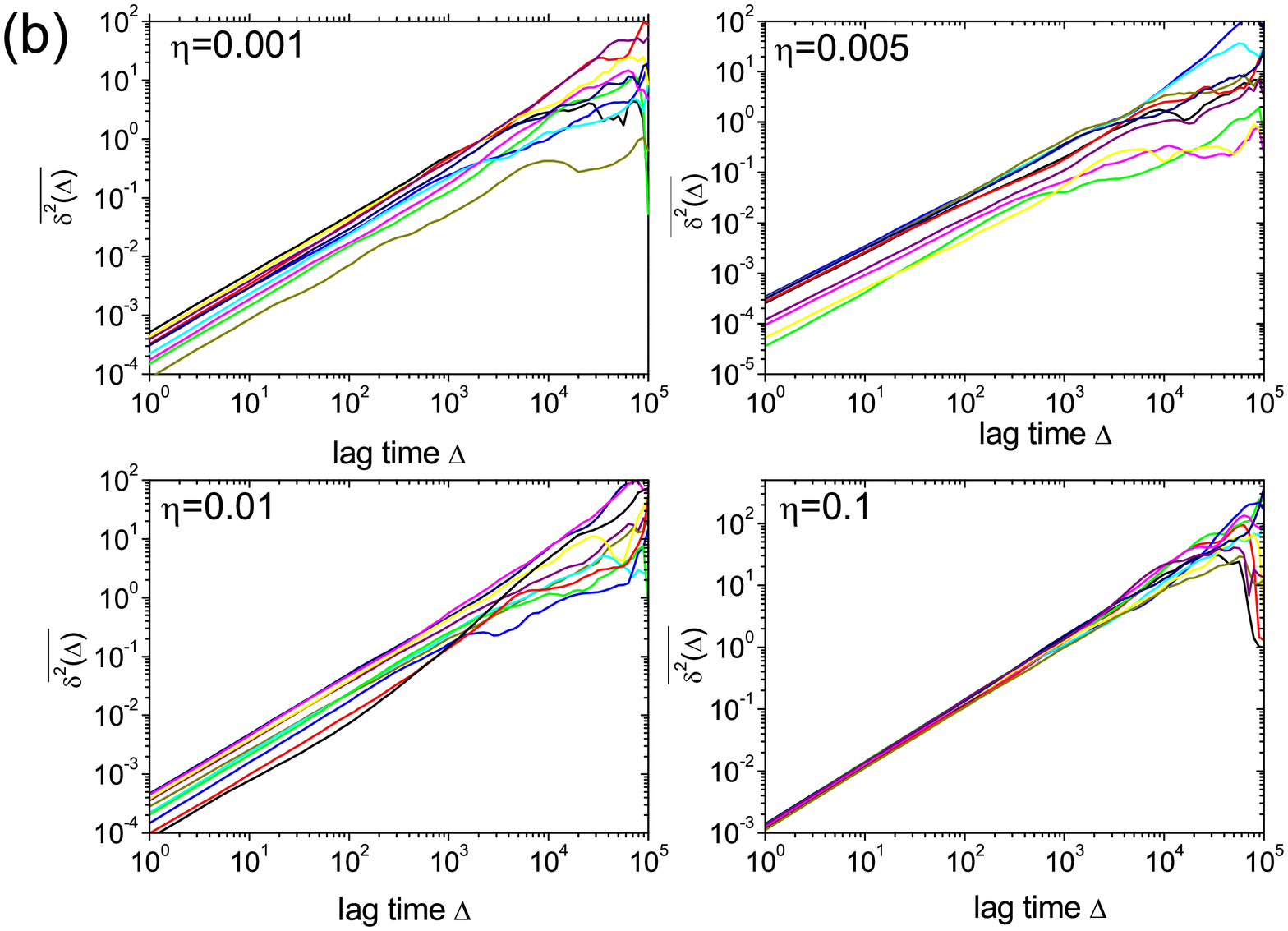}
\caption{Ten individual time averaged MSD curves of the nCTRW process with
Ornstein-Uhlenbeck noise of strengths $\eta=0.001$, $0.005$, $0.01$, and $0.1$
for anomalous diffusion exponents (a) $\alpha=0.5$ and (b) $\alpha=0.8$. 
$T=10^5$ is used as in Fig. \ref{fig:outamsd}.}
\label{outamsds}
\end{figure*}

Interestingly, the time averaged MSD for the nCTRW with anomalous diffusion
exponent $\alpha=0.5$ and noise strength $\eta=0.1$ is reminiscent of the
MSD curves observed for micron-sized tracer particles immersed in wormlike
micellar solutions, which are known to behave as a viscoelastic polymer network
when the micelles concentration is above a critical value \cite{dreiss}.
Experiments using diffusing wave spectroscopy \cite{galvan,bellour} and
single-particle tracking \cite{natascha1} revealed that the immersed particles
exhibit three distinct diffusive behaviors in different time windows. It was
shown that particles surrounded by the micellar network undergo a Brownian
diffusion at short (sub-milliseconds) times until they engage with the caging
effects of the micellar network, while at later times (milliseconds to
sub-seconds) one observes a seemingly confined Brownian motion. It
turns out that this confined diffusion is in fact a pronounced subdiffusive
motion characterized by anti-persistent spatial correlation induced by
the polymer network, governed by the fractional Langevin equation
\cite{natascha1}. At macroscopic times, when the wormlike micelle solution
behaves as a viscous fluid, the particle again shows a Brownian diffusion,
albeit with a significantly reduced diffusion constant. The results obtained
here suggest that due to the presence of the Ornstein-Uhlenbeck noise the
CTRW process could be mistakenly interpreted to conform to a physically
different system, i.e., Brownian motion, and, thus, one needs to be careful
in analyzing the data with several possible models, and to use several
complementary diagnosis tools.

The expression \eqref{outamsd} for the time averaged MSD suggests that
$\overline{\delta^2(\Delta,T)}$ stops aging and reflects almost entirely
the character of the Ornstein-Uhlenbeck noise process if $T\gtrsim T_{\mathrm{
cr}}\sim(kK_\alpha \Delta/[\Gamma(1+\alpha)\eta^2D])^{1/(1-\alpha)}$. This
is indeed shown in Fig.~\ref{fig:agingou} where the time averaged MSD
$\langle\overline{\delta^2(\Delta,T)}\rangle$ is plotted as a function
of the overall measurement time $T$ at a fixed lag time $\Delta=100$
for nCTRW with $\alpha=0.5$, together with the theoretical prediction
Eq.~\eqref{outamsd}. For the weakest noise $\eta=0.001$ whose crossover
time $T_{\mathrm{cr}}\sim10^9$ is beyond $T_{\mathrm{max}}=10^7$, the time
averaged MSD only displays aging of the naked CTRW process with scaling $\sim
T^{\alpha-1}$. When the noise strength is increased to $\eta=0.01$, the
time averaged MSD starts to show ergodic behavior as the measurement time $T$
gets larger than the crossover time $T_{\mathrm{cr}}\sim10^5$. In the extreme
case when the nCTRW process is dominated by the noise (here $\eta=0.1$),
effectively no aging is observed in $\overline{\delta^2(\Delta,T)}$ and the
process appears ergodic.

In Fig.~\ref{outamsds} we plot ten individual time averaged MSD curves
$\overline{\delta^2(\Delta)}$. For the case of more pronounced
subdiffusion ($\alpha=0.5$), the trajectory-to-trajectory variations are
significant, due to the combined effect of long-time stalling events and the
Ornstein-Uhlenbeck noise. Thus, for the smallest noise strength $\eta=0.001$
some time averaged MSDs exhibit a large deviation from the linear scaling $\sim
\Delta$ expected from the trajectory-averaged time averaged MSD. In this case,
the long stalling events, that are of the order of the measurement time $T$,
lead to the plateaus in $\overline{\delta^2(\Delta)}$. Intriguingly, such
plateaus also appear in the presence of the largest noise strength, $\eta=0.1$.
Here, however, they represent the confined diffusion of the Ornstein-Uhlenbeck
noise in which one or few long-stalling events occur in the the CTRW process.
For the CTRW process with more frequent jumps ($\alpha=0.8$) the individual
time averaged MSDs follow the expected scaling behavior with smaller amplitude
fluctuations in $\overline{\delta^2(\Delta)}$, as now the noise is relatively
stronger.

\subsection{Scatter distribution}

\begin{figure}
\begin{center}
\includegraphics[width=8.8cm,angle=0]{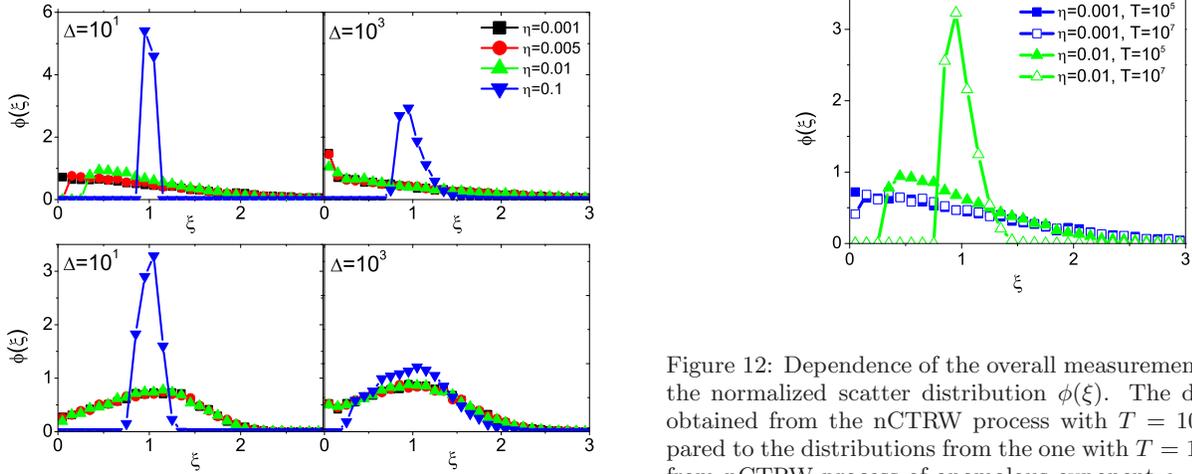}
\caption{Normalized scatter distribution $\phi(\xi)$ as a
function of the dimensionless variable $\xi=\overline{\delta^2}/\langle
\overline{\delta^2}\rangle$ for the cases of $\eta=0.001$ (black square),
0.005 (red circle), 0.01 (green upper-triangle), and 0.1 (blue down-triangle).
Top: $\alpha=0.5$. Bottom: $\alpha=0.8$. In each panel the results were
obtained from $10^4$ runs.}
\label{ouscatter}
\end{center}
\end{figure}

Fig.~\ref{ouscatter} illustrates the normalized scatter distribution $\phi(\xi)$
of the individual time averaged MSDs obtained from $10^4$ trajectories, as a
function of the dimensionless variable $\xi=\overline{\delta^2}/\langle
\overline {\delta^2}\rangle$. The overall behavior is qualitatively consistent
with those of the scatter distributions for added Brownian noise, as shown in
Fig.~\ref{bmscatter}. The distribution becomes increasingly ergodic as the
Gaussian noise increases, especially for the case of the more pronounced
subdiffusive process ($\alpha=0.5$). Here, the finite contribution at $\xi=0$
in $\phi(\xi)$ is gradually suppressed and the peak of the distributions is
approaching $\xi=1$ as the strength of the noise is increased.

\begin{figure}
\centering
\includegraphics[width=6cm,angle=0]{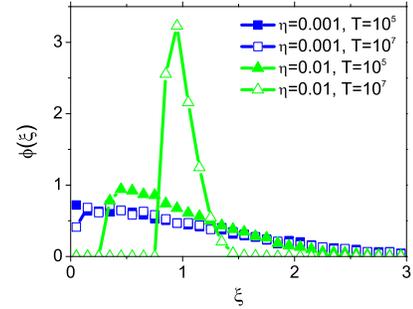}
\caption{Dependence of the overall measurement time $T$ on the normalized
scatter distribution $\phi(\xi)$. The distributions obtained from the nCTRW
process with $T=10^7$ are compared to the distributions from the one with
$T=10^5$. Results from nCTRW process of anomalous exponent $\alpha=0.5$
mixed with $\eta z_{OU}(t)$ at $\eta=0.001$ and 0.01.}
\label{scatterouT}
\end{figure}

From the time averaged MSD \eqref{outamsd} and its aging property in
Fig.~\ref{fig:agingou}
it is also expected that the distribution attains apparent ergodic features as
the observation time $T$ is increased. We show this in Fig.~\ref{scatterouT}
where the scatter distributions for the same nCTRW process simulated up to
$T=10^7$ are compared to the previous ones with $T=10^5$. The general trend is
that $\phi(\xi)$ becomes sharper as $T$ increases. Importantly, this ergodic
effect becomes relevant provided that $T$ is increased to at least be comparable
with the crossover time $T_{\mathrm{cr}}\sim(kK_\alpha\Delta/[\Gamma(1+\alpha)
\eta^2D])^{1/(1-
\alpha)}$ defined above. As seen for the case for $\eta=0.001$, the distribution
$\phi(\xi)$ is almost unaffected, except at $\xi\approx0$ where $T\ll T_{
\mathrm{cr}}(\sim 10^9)$. In contrast to this, the distribution at $\eta=0.01$
is noticeably narrower around $\xi=1$ when $T$ is increased ($T_{\mathrm{cr}}
\sim10^5$).

\subsection{$p$-variation test}

\begin{figure*}
\centering
\includegraphics[width=12cm,angle=0]{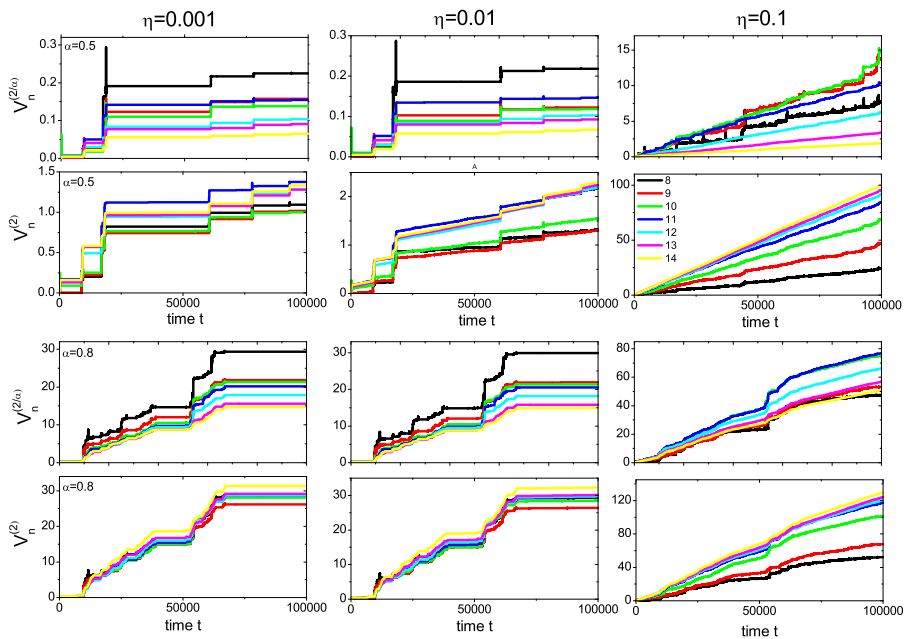} 
\caption{Results of the $p$-variation test for nCTRW with superimposed
Ornstein-Uhlenbeck noise $\eta z_{OU}(t)$ for noise strengths $\eta=
0.001$, $\eta=0.01$, and $\eta=0.1$. The upper and lower two rows
are for $\alpha=0.5$ and $0.8$. Same color codes as those
of Fig.~\ref{pvar1}.}
\label{pvar2}
\end{figure*}

In Fig.~\ref{pvar2} we show the variation of the $p$-sums $V_n^{(2/\alpha)}(t)$
and $V_n^{(2)}(t)$ at increasing $n$ for the simulated trajectories $x(t)$ of
Fig.~\ref{outra}. For the smallest noise strength $\eta=0.001$ shown in
Fig.~\ref{pvar2} the $p$-variation test produces the result expected for
the naked CTRW process. In the case of nCTRW with $\alpha=0.8$ this behavior
persists in the presence of large noise strengths up to $\eta=0.01$. This is
due to the fact that the naked CTRW process $x_\alpha(t)$ features large
displacements relative to those of the Ornstein-Uhlenbeck process. In contrast,
for $\alpha=0.5$ the profiles of $V_n^{(2)}(t)$ are mildly affected by the
Ornstein-Uhlenbeck noise. Namely, the plateaus are tilted, with a somewhat
larger slope at higher $\eta$, while their overall profiles preserve features
of the monotonic, step-like behavior of the CTRW process $x_\alpha(t)$.

Concurrently, we note that at larger values of $\eta$ neither $V_n^{(2)}(t)$ nor
$V_n^{(2/\alpha)}(t)$ show any indication of CTRW for both $\alpha=0.5$ and $0.8$.
At these noise strengths, the Ornstein-Uhlenbeck noise dominates the
$p$-variation results. As explained in Section \ref{analyses} and shown in
Fig.~\ref{pvar}, in the limit $n\rightarrow\infty$ the $p$-sums for the
Ornstein-Uhlenbeck noise converge to the results for Brownian noise. Thus,
while $V_n^{(2/\alpha)}(t)$ rapidly decreases with increasing $n$ in the case of
Brownian noise, such a tendency is also present for the Ornstein-Uhlenbeck noise.
Moreover, the spike-like profiles in $V_n^{(4)}(t)$ at $n=8$ for $\eta=0.1$
reflects the property of the Ornstein-Uhlenbeck noise (see Fig.~\ref{pvar}).
For $p=2$, we find that the $p$-sum of the Ornstein-Uhlenbeck noise$z_{OU}(t)$
behaves as $V_n^{(2)}(t)\sim\frac{2D}{k}\left[
1-\exp(-\frac{k}{2}\frac{T}{2^n})\right]\left(\frac{2^n}{T}\right)t$ (see
App.~\ref{app_pvar}). This result explains why $V_n^{(2)}(t)$  monotonically
increases with $n$ up to $n\sim\log(kT)/\log2\approx10$, and for $n$ larger
it grows as $V_n^{(2)}(t)\sim 2Dt$, as shown in Fig. \ref{pvar2}. As in the case
of the above nCTRW in the presence of Brownian noise $z_B(t)$, we find that
the $p$-variation result may be substantially affected by the added
Ornstein-Uhlenbeck noise, and the identification of the underlying CTRW process
become impossible.

\section{Conclusions}
\label{conc}

We introduced the noisy CTRW process, in which an ordinary CTRW process is
superimposed with Gaussian noise, representing physically relevant cases
when the pure CTRW motion becomes distorted by a noisy environment. We
investigated how the additional ergodic noise interferes with the non-ergodic
behaviors of the underlying subdiffusive CTRW motion. Considering the two
types of Gaussian noise, Brownian noise and Ornstein-Uhlenbeck noise,
we simulated the resulting nCTRW motion and studied physical quantities
such as the ensemble and time averaged MSDs, the amplitude scatter
distribution, and the behavior of the $p$-variation.

The analysis demonstrates that the influence of the Gaussian noise on these
statistical quantities is highly specific to the quantity of concern. Moreover,
it depends not only on the type of the noise and its strength but also on
the length of the trajectory and the time scale. Depending on those specific
conditions a quantitative analysis of the nCTRW process may reveal or mask the
underlying non-ergodicity of the naked CTRW process. Thus care is needed when
we want to diagnose the stochastic nature of a physical process based on
experimental data. One way to avoid wrong conclusions is to apply complementary
analysis techniques, such as the quantities used herein, or moment ratios,
mean maximal excursion methods, first passage dynamics, or others. The other
necessary ingredient is a good physical intuition for the observed process. It
would be also interesting to find analytical expressions for the $p$-variations,
of mixed processes, as those we have considered here. Our simulations results
show that on finite measurement time the noise is crucial, and could easily
destroy our basic understanding of the underlying process.

From the present study an experimentally relevant inverse problem can be
posed. Can one filter out the Gaussian noise from the experimentally obtained
nCTRW process? Although obtaining the noise-cleansed profile from a given nCTRW
trajectory may appear infeasible, one could in principle obtain noise-free
contribution in some ensemble- or time-averaged physical quantities of
the nCTRW process provided one is able to attain a sufficiently long trajectory.
In the ensemble averaged MSD the noise survives if the noise is Brownian
motion while the CTRW process wins if the noise is of Ornstein-Uhlenbeck type
[see Eq.~\eqref{oumsd}]. This is of course what we expect, since the MSD
of the Brownian motion increases linearly with time, for CTRW like $t^\alpha$,
and is a constant for Ornstein-Uhlenbeck---hence it is not surprising to see
this behavior. In contrast for the time averaged MSD the dominant contribution
comes always from the noise [Eqs.~(\ref{bmeatamsd}) and (\ref{outamsd})] in the
sense that when the measurement time $T$ is very large even the bounded
Ornstein-Uhlenbeck noise wins over. This is due to an aging effect, the time
averaged MSD of the CTRW process decreases with measurement time $T$. We thus
see that the influence of the noise on time averages is fundamentally
different from that on ensemble averages.
As an example, one can extract almost solely the noise contribution in the 
time-averaged MSD (and not the bare CTRW itself) from a
very long nCTRW trajectory. For an ergodic process this corresponds to the almost
identical noise contribution in the ensemble-averaged MSD, namely, 
$\eta^2\langle z^2(\Delta) \rangle\simeq \langle\overline{\delta^2(\Delta,T\rightarrow\infty)}\rangle$.
How to subtract this noise from real data is left for future work.

The nCTRW process developed herein is a physical extension of pure CTRW
dynamics. We believe that it represents an important advance in the truthful
description of anomalous diffusion data in thermal microscopic systems, where
the environment is noisy by definition. Mathematically, the nCTRW process is
quite intuitive, due to the additivity of the Gaussian noise.

Our present study can naturally be extended to more complicated noise sources,
such as fractional Brownian motion (FBM), in order to obtain insight into
intracellular anomalous diffusion that show both CTRW and FBM behaviors. It is
expected that although FBM-like noise should lead to similar effects on the
statistical behavior of the nCTRW process, the quantitative results will be
profoundly different due to the scaling law of the FBM-like noise $\langle
z_{\mathrm{FBM}}^2(t)\rangle\sim t^{\alpha'}$ with exponent $0<\alpha'<2$.

\acknowledgments

We acknowledge funding from the Academy of Finland within the Finland
Distinguished Professor (FiDiPro) scheme and from the Israel Science Foundation.

\begin{appendix}

\section{$p$-variation of CTRW subdiffusion and Ornstein-Uhlenbeck noise}
\label{app_pvar}

We here discuss the $p$-variation properties of CTRW subdiffusion and the
Ornstein-Uhlenbeck noise in some more detail.

\emph{CTRW subdiffusion.}
The subdiffusive CTRW process $x_\alpha(t)$ with its PDF governed by the
fractional Fokker-Planck equation (\ref{FFPE}) can be described through
the subordinated Brownian motion $x_\alpha(t)=B(S_\alpha(t))$, where $B(\tau)$
defined with the internal time $\tau$ is an ordinary Brownian motion
satisfying $\langle B(\tau)\rangle=0$ and $\langle B^2(\tau)\rangle=2D\tau$.
Here, $S_\alpha(t)$ is the so called inverse subordinator matching the
laboratory time $t$ to the internal time $\tau$. For the CTRW process
$x_\alpha(t)$, the $p$-sum $V_n^{(2)}$(t) as $n$ grows to infinity satisfies
\cite{marcin}
\begin{equation}
V^{(2)}(t)= 2DS_\alpha(t).
\end{equation}
As shown in Fig. 1, $S_\alpha(t)$ has a step-like incremental profile and its
jump times represent those for a given realization of CTRW process
$x_\alpha(t)$. On the other hand, the $p$-sum $V_n^{(2/\alpha)}(t)$ (with
$0<\alpha<1$) decreases with increasing $n$, finally $V^{(2/\alpha)}(t)=0$ at
$n\rightarrow\infty$ \cite{marcin}. This is also shown in the simulations in
Fig.~\ref{pvar}.

\emph{Ornstein-Uhlenbeck noise.} The ensemble average of the $p$-variation
sum at finite $n$ becomes
\begin{equation}
\left<V_n^{(2)}(t)\right>\approx \frac{2D}{k}\left[1-\exp\left(-\frac{k}{2}
\frac{T}{2^n}\right)\right]\left(\frac{2^n}{T}\right)t,
\end{equation}
neglecting an additional term which becomes negligible for large $T$.
For small $n\lesssim \log(kT)/\log2$, the above $p$ sum simplifies to
\begin{equation}
\langle V_n^{(2)}(t)\rangle\approx\frac{2D}{k}\frac{2^n}{T}t.
\end{equation}
In this case the linear slope of $\langle V^{(2)}(t)\rangle$ increased with
$n$ up to values $\log(kT)/\log2\sim 10$ (for the given parameter values used
in our simulation). This is shown in Fig.~\ref{pvar}. In the other case, when
$n\rightarrow\infty$, the $p$ sum converges to the result of the Brownian
noise $\langle V^{(2)}(t)\rangle\approx 2Dt$. Hence, for large $n\gtrsim
\log(kT)/\log2$, $\langle V^{(2)}(t)\rangle$ is proportional to $t$ with an
$n$-independent slope. From the fact that on short-time scales (as $n\to\infty$)
the Ornstein-Uhlenbeck process behaves like a free Brownian process, it can be
inferred that its $p$-variation results are identical with those for simple
Brownian motion. Therefore, $V^{(4)}(t)$ is expected to converge to zero as
$n$ increases. This is indeed observed in the simulations result in
Fig.~\ref{pvar}. We find that when $n$ is small, the increment of $V^{(4)}
(t)$ exhibits a spike-like profile. This behavior presumably occurs due to the
fact that the quartic moment of the displacement $x((j+1)T/[2^n])-x(jT/[2
^n])$ becomes very small when the lag time $T/2^n$ is larger than the
relaxation time $1/k$ of the process for small $n$.

\end{appendix}

\end{document}